\documentclass[11pt,letterpaper]{article}
\pdfoutput=1

\usepackage{jcappub}
\usepackage{amsmath,amssymb,color}
\usepackage{enumerate,xstring}
\usepackage{jcappub}
\usepackage{amsmath}
\usepackage{amsfonts}
\usepackage{amssymb}
\usepackage{graphicx}
\usepackage{epsfig}
\usepackage{color}
\usepackage{multirow}
\usepackage{graphicx}
\usepackage{bigints}
\usepackage{todonotes,bm,etoolbox}
\usepackage{calligra}
\usepackage{hyperref}
\usepackage{tabularx}
\usepackage{booktabs}

\usepackage{tikz}

\def\ba#1\ea{\begin{align}#1\end{align}}
\def\bea{\begin{eqnarray}}
\def\eea{\end{eqnarray}}
\def\be{\begin{equation}}
\def\ee{\end{equation}}

\def\({\left(}
\def\){\right)}
\def\[{\left[}
\def\]{\right]}

\def\<{\left\langle}
\def\>{\right\rangle}

\def\comment#1{}

\def\eps{\epsilon}

\renewcommand{\v}[1]{\bm{#1}}
\renewcommand{\vec}[1]{\bm{#1}}

\def\vx{\v{x}}
\def\vk{\v{k}}

\newcommand{\perm}[1]{ \expandafter\ifstrempty\expandafter{#1} {\mbox{perm.}} {\mbox{$#1$ perm.}} }

\def\O{\mathcal{O}}

\def\R{\mathcal{R}} % \zeta

\newcommand{\fnl}{f_\textnormal{\textsc{nl}}}
\newcommand{\acip}{A}

\definecolor{RedWine}{rgb}{0.743,0,0}
\definecolor{RoyalBlue}{rgb}{0.25,.41,.88}
\definecolor{ForestGreen}{rgb}{.13,.54,.13}
\definecolor{Goldenrod}{rgb}{.85,.65,.13}

\newcommand{\bq}{\begin{eqnarray}}
\newcommand{\eq}{\end{eqnarray}}

\title{\huge Compensated Isocurvature Perturbations in the Galaxy Power Spectrum}

\author[a]{Alexandre Barreira,}
\author[a]{Giovanni Cabass,}
\author[a]{Kaloian D. Lozanov,}
\author[a]{and Fabian~Schmidt}

\affiliation[a]{
Max-Planck-Institut f\"ur Astrophysik, Karl-Schwarzschild-Stra\ss e~1, 85748 Garching, Germany
}

\emailAdd{barreira@mpa-garching.mpg.de}
\emailAdd{gcabass@mpa-garching.mpg.de}
\emailAdd{klozanov@mpa-garching.mpg.de}
\emailAdd{fabians@mpa-garching.mpg.de}

\date{\today}

\abstract{We investigate the potential of the galaxy power spectrum to constrain compensated 
isocurvature perturbations (CIPs), primordial fluctuations in the baryon density that are compensated by fluctuations in CDM density to ensure an unperturbed total matter density. We show that CIPs contribute to the galaxy overdensity at linear order, and if they are close to scale-invariant,
their effects are nearly perfectly degenerate with the local PNG parameter 
$\fnl$ if they correlate with the adiabatic perturbations. 
This degeneracy can however be broken by analyzing multiple galaxy samples with 
different bias parameters, or by taking CMB priors on $\fnl$ into account. 
Parametrizing the amplitude of the CIP power spectrum as $P_{\sigma\sigma} = {\acip}^2P_{\R\R}$ 
(where $P_{\R\R}$ is the adiabatic power spectrum) we find, for a number of fiducial galaxy samples in a simplified forecast setup, 
that constraints on $\acip$, relative to those on $\fnl$, of order $\sigma_{\acip}/\sigma_{\fnl} \approx 1-2$ are achievable for CIPs correlated with adiabatic perturbations, and 
$\sigma_{\acip}/\sigma_{\fnl} \approx 5$ for the uncorrelated case. These values are independent of survey volume, and suggest that current galaxy data are already able to 
improve significantly on the tightest existing constraints on CIPs from the CMB. Future galaxy surveys that aim to achieve $\sigma_{\fnl} \sim 1$ have the potential to place even stronger bounds on CIPs.} 

\begin{document}

\maketitle

%**********************
%**********************
\section{Introduction}
\label{sec:introduction}

Understanding the statistical properties of the primordial density fluctuations of our Universe 
is one of the current main goals in theoretical and observational cosmology. 
The current tightest constraints come from analyses of the cosmic microwave background (CMB) 
data obtained by the Planck satellite \cite{2018arXiv180706211P, 2019arXiv190505697P, 2018arXiv180706209P}, 
and they are compatible with Gaussian adiabatic fluctuations as predicted by the simplest single-field models of inflation. That is, 
the fluctuations in energy of each species are all given in terms of Gaussian-distributed curvature perturbations $\R(\vx)$. 
Any detected departure from adiabaticity and Gaussianity would have immediate and important ramifications 
on our knowledge of the physics of the early universe that generated the fluctuations, 
as well as the physics of the late-time universe that evolved out of them.

For example, if more than one field is present during the epoch of inflation, there will generically also be isocurvature 
perturbations in addition to the adiabatic ones, i.e.~relative density perturbations between the different matter and radiation species. 
Isocurvature perturbations are typically defined with respect to the photon number density as 
$S_{i\gamma} = \delta n_i/\bar{n}_i - \delta n_\gamma/\bar{n}_\gamma$, where $\bar{n}_i$ and $\delta n_{i}$ 
are the mean particle number density and its fluctuation for species $i$, respectively, while $\gamma$ denotes photons. 
These isocurvature modes are well constrained by the Planck CMB data, with 
the latest analysis constraining their contribution to the total 
CMB power spectrum to be less than $2\%$ \cite{2018arXiv180706211P}.

Interestingly, however, there is a mode of isocurvature perturbations that largely escapes 
the constraining power of the CMB data; these are called compensated isocurvature perturbations (CIP) 
\cite{2003PhRvD..67l3513G, 2009PhRvD..80f3535G, 2010ApJ...716..907H, grin/dore/kamionkowski, 
2014PhRvD..89b3006G, 2016PhRvD..94d3534H, 2016PhRvD..93d3008M, 2017PhRvD..96h3508S, 2019MNRAS.485.1248S, 2019arXiv190400024H}. 
A CIP, which we denote in this paper by $\sigma(\vx)$, is characterized 
by $S_{c\gamma} = - (\Omega_b/\Omega_c)S_{b\gamma}$, i.e.~fluctuations in the baryons ($b$) 
that are compensated by fluctuations in the cold dark matter (CDM, $c$). When baryons are nonrelativistic, these leave the total matter ($m$) distribution unchanged and adiabatic, 
$S_{m\gamma} = (\Omega_c/\Omega_m)S_{c\gamma} + (\Omega_b/\Omega_m)S_{b\gamma} = 0$.
This means that during matter domination, gravitational potentials are unchanged at linear order. 
For the phenomenology of the CMB then, CIPs only induce a spatial modulation 
of the photon-to-baryon ratio, and therefore of the plasma sound speed, which results in a smoothing of the higher 
power spectrum multipoles and induces a specific connected four-point function (trispectrum) in the observed CMB anisotropies 
\cite{grin/dore/kamionkowski, 2014PhRvD..89b3006G, 2016PhRvD..93d3008M, 2017JCAP...04..014V, 2017PhRvD..96h3508S}. 
These are only second-order effects, hence CIPs end up being currently rather poorly constrained by the CMB data: 
some recent constraint studies \cite{2018arXiv180706211P, 2017PhRvD..96h3508S, 2016PhRvD..93d3008M, 2017JCAP...04..014V} 
still allow for amplitudes of the primordial CIP power spectrum to be over $5$ orders of magnitude larger 
than the amplitude of the adiabatic power spectrum! This remarkably loose constraint on CIPs indicates that there is still 
much room for progress to be made in our understanding of early-universe physics. While isocurvature perturbations cannot be generated during inflation if the energy density is dominated by a single scalar field, 
any realistic model must feature multi-field dynamics (e.g.~to allow for the reheating phase). This could lead to the production of isocurvature, including CIPs. Indeed, CIPs can be generated in multi-field models of inflation like the curvaton scenario 
\cite{1997PhRvD..56..535L, 2000PhRvD..62d3504L, 2003PhRvD..67l3513G, 2003PhRvD..67b3503L, 2006RvMP...78..537B, 2006PhRvD..74j3003S, 2015PhRvD..92f3018H}, 
as well as in baryogenesis scenarios driven by a scalar field \cite{2016JCAP...08..052D}.

The late-time large-scale structure in the universe
can also be used to constrain CIPs by inspecting observables that would be sensitive to spatial modulations 
of the relative abundance of baryons and CDM. Examples include studies of baryon/gas fractions in galaxy clusters 
\cite{2010ApJ...716..907H}, mass-weighted vs.~luminosity-weighted galaxy statistics \cite{2019MNRAS.485.1248S}, $21\rm cm$ 
line intensity mapping \cite{2009PhRvD..80f3535G} and spatial modulations of the baryon acoustic oscillation (BAO) 
features imprinted in the galaxy distribution \cite{2019arXiv190400024H}. 
More recently, Ref.~\cite{2020JCAP...02..005B} showed, using cosmological simulations, 
that galaxy formation and evolution is sensitive to the presence of long-wavelength CIPs. 
Specifically, galaxies originating from regions inside a CIP with an excess of baryon density exist 
in fewer numbers when selected in terms of total host halo mass: 
this is because of the impact of an excess of baryons (at fixed total matter density) 
on the shape of the matter power spectrum after recombination. 
On the other hand, the same galaxies will have more of their total mass in stars, 
which can in fact revert the trend and yield an excess of galaxy numbers if selected by stellar mass. 
This demonstrated sensitivity of the galaxy abundance to primordial CIPs opens the possibility 
to use the statistics of their distribution to constrain the amplitude of the power spectrum of primordial CIPs \cite{2019arXiv190808953H}.

Another common consequence of models beyond single-field inflation is primordial non-Gaus{\-}sianity (PNG) of the local type.
The latter is popularly parametrized by the parameter $\fnl$ \cite{2001PhRvD..63f3002K} as 
$\phi(\vx) = \phi_{\rm G}(\vx) + \fnl\left[\phi_{\rm G}^2(\vx) - \langle\phi_{\rm G}^2(\vx)\rangle\right]$, 
where $\phi(\vx) = (3/5)\R(\vx)$ is the primordial Bardeen gravitational
potential after inflation, $\phi_{\rm G}(\vx)$ is a Gaussian-distributed field and $\langle\dots\rangle$ denotes the ensemble average. 
We will see that this is relevant in studies of CIPs using the galaxy distribution because, if CIPs correlate with the adiabatic perturbations, then they contribute to the galaxy power spectrum 
with the same scale dependence as local PNG (this is often called the 
{\it scale-dependent bias} contribution \cite{dalal/etal:2008}). 
Constraints from current galaxy surveys typically yield error bars on $\fnl$ of order $\sigma_{\fnl} \sim 50$ 
\cite{slosar/etal:2008, 2013MNRAS.428.1116R, 2014PhRvD..89b3511G, 2014PhRvL.113v1301L, 2014MNRAS.441L..16G, 2015JCAP...05..040H, 2019JCAP...09..010C} and future galaxy surveys aim to 
bring this down to $\sigma_{\fnl} \sim 1$ \cite{2012MNRAS.422.2854G, 2014arXiv1412.4872D, 2014arXiv1412.4671A, 2015JCAP...01..042R, 2015PhRvD..92f3525A, 2015MNRAS.448.1035C, 2017PhRvD..95l3513D, 2017PDU....15...35R}. 
If CIPs and local PNG contribute similarly to the galaxy distribution, it follows that experiments 
aiming to place tight constraints on $\fnl$ should be able to place equally tight constraints on primordial CIPs. 
In fact, Ref.~\cite{2019arXiv190808953H} recently confirmed this in a forecast study of the constraining power on correlated CIPs 
from the cross-correlation of galaxies with the kinetic Sunyaev-Zel'dovich (kSZ) effect. 

In this paper we investigate the potential of using the galaxy distribution alone to constrain CIPs. We will focus on analyses of the galaxy power spectrum and we will see that with a single galaxy sample one is effectively unable to simultaneously constrain both PNG and (correlated) CIPs due to a strong degeneracy that exists between their effects. Notice that this means that all current constraints on $\fnl$ from galaxy clustering are heavily influenced by the (usually implicit) assumption of zero CIP, a fact which has not been recognized widely before. The degeneracy can however be broken efficiently with the galaxy multitracer technique \cite{2009JCAP...10..007M, 2009PhRvL.102b1302S}, in which a given galaxy sample is split into at least two with different bias properties, and the corresponding auto- and cross-spectra are analysed. Our main goal here is not to draw precise quantitative statements on the constraints that can be obtained with specific surveys, but rather to illustrate and discuss some aspects of the phenomenology behind such analyses. Our numerical results will show, nonetheless, that the galaxy power spectrum can indeed be one of the strongest probes of primordial CIPs, with a constraining power that is similar to that attained in local PNG constraints/forecasts; this in fact leaves open the possibility for existing galaxy samples to be able to beat already the tightest current bounds from the CMB.

The rest of this paper is organized as follows. In Sec.~\ref{sec:model} we describe and discuss the contributions from 
local PNG and CIPs to the galaxy power spectrum and its covariance. 
Section~\ref{sec:results} contains our main numerical results and findings, 
where we discuss in particular the important effects of galaxy bias (Sec.~\ref{sec:results:bias}), 
CIP correlation with the adiabatic mode (Sec.~\ref{sec:results:xi}) 
and prior information on $\fnl$ (Sec.~\ref{sec:results:cmbprior}) on the resulting local PNG and CIP constraints. 
We also compare the constraining power of galaxy power spectra analyses with that of existing constraints 
and forecasts on CIPs in Sec.~\ref{sec:results:comparison}, and comment on the impact of a number of simplified aspects of our analysis in Sec.~\ref{sec:results:discussion}.
Appendix \ref{app:estimator} contains more details on the derivation of the covariance matrix 
and calculation of the galaxy power spectrum data vector. 
Throughout this paper, we work with a fiducial standard flat-$\Lambda{\rm CDM}$ cosmology with physical cosmic baryon 
density parameter $\Omega_{b0}h^2 = 0.02119$, physical cosmic CDM density parameter $\Omega_{c0}h^2 = 0.1206$, 
Hubble rate today $H_0 = 100 h\ {\rm km/s/Mpc}$ with $h = 0.6774$, and spectral index of the primordial adiabatic 
scalar power spectrum $n_{\rm s} = 0.9667$ with an amplitude $A_{\rm s} = 2.068\times10^{-9}$ 
at a pivot scale $k_{\rm p} = 0.05\ {\rm Mpc}^{-1}$. We set the fiducial amplitudes of local PNG and primordial CIPs to zero. Throughout, we ignore the effects of massive neutrinos on the growth of structure, since the effects studied here are prominent on scales much larger than the neutrino free-streaming scale.

%**********************
%**********************
\section{Galaxy clustering with CIPs and local PNG}
\label{sec:model}

In this section we describe the calculation of the galaxy power spectrum (and corresponding covariance) 
that we consider to investigate the prospects of galaxy clustering analyses to constrain compensated isocurvature perturbations.

%**********************
%**********************
\subsection{The galaxy power spectrum}
\label{sec:model:pk}

\begin{figure}[t!]
	\centering
          \includegraphics[width=9cm]{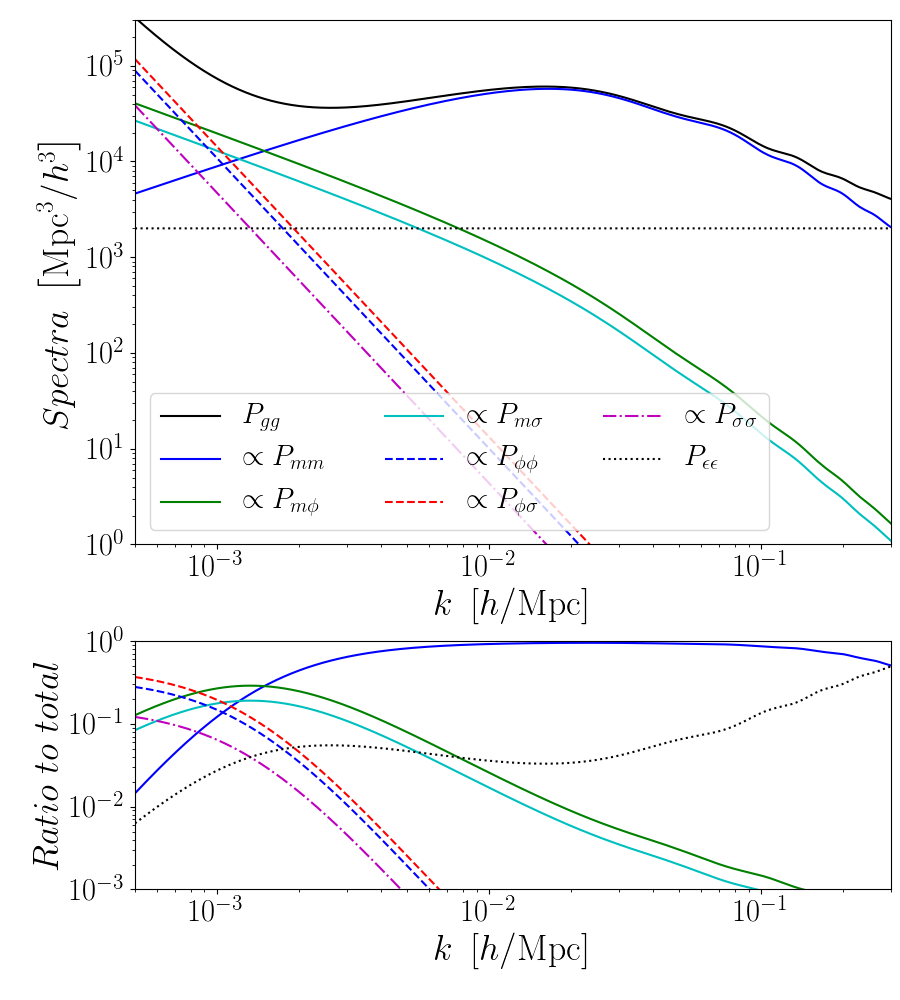}
	\caption{Contributions from local PNG and primordial CIPs to the galaxy power spectrum. 
	The upper panel shows the prediction from all the terms contributing to Eq.~(\ref{eq:Pgg}), as labeled. 
	The lower panel shows the ratio of each contribution to the total. 
	The result shown corresponds to $z=1$, $b_1 = 2.5$, $b_{\phi} = 5.058$, $b_{\sigma} = 0.5$, 
	$\fnl = 5$, $\acip = 20$, $\xi = 1$ and $\bar{n}_g = 5\times10^{-4}\ h^3/{\rm Mpc}^3$.
        } 
\label{fig:model}
\end{figure}

Assuming the existence of local-type PNG in addition to CIP, we begin with the general, linear-order expression for the rest-frame galaxy density contrast at position $\vx$ and redshift $z$, $\delta_g(\vx, z)$:\footnote{Strictly speaking, $\phi$ and $\sigma$ should be evaluated at the Lagrangian position corresponding to the Eulerian position $\vx$ \cite{assassi/baumann/schmidt}, but this distinction is only relevant at second order.}
\bq\label{eq:biasexp}
\delta_g(\vx, z) = b_1(z) \delta_m(\vx, z) + b_{\rm \phi}(z) \fnl \phi(\vx) + b_{\sigma}(z) \sigma(\vx) + \eps(\vx)\,\,,
\eq
where $\delta_m$ is the total matter density contrast, $\phi$ is the primordial Bardeen potential generated by inflation, 
$\sigma$ is the primordial compensated baryon-CDM isocurvature perturbation and $\eps$ is a stochastic contribution 
to the galaxy overdensity. In our convention, a positive CIP $\sigma > 0$ corresponds to an excess of baryon density, 
$\smash{\rho_{b}(\vx, z) = \bar{\rho}_b(z)\left[1 + \sigma(\vx)\right]}$, 
and compensated suppression of CDM density $\smash{\rho_{c}(\vx, z) = \bar{\rho}_c(z)\left[1 - f_b\sigma(\vx)\right]}$, 
where $f_b = \Omega_{b0}/\Omega_{c0}$ is the ratio of the cosmic baryon to CDM density. The bias terms $b_1$, $b_\phi$ and $b_\sigma$ are functions of redshift and of the properties of the galaxies, and they effectively encode how galaxy formation changes (or {\it responds}) to the presence of long-wavelength $\delta_m$,  $\phi$ and $\sigma$ perturbations, respectively (see Ref.~\cite{biasreview} for a review on galaxy bias). Processes such as baryonic accretion, star formation and black hole growth/feedback are sensitive to the local amount of baryons (even for zero total matter fluctuations), which is why CIPs contribute to linear order in Eq.~(\ref{eq:biasexp}). For example, at fixed star formation efficiency, the number of stars formed is directly proportional to the baryon density. Further, note that large-scale primordial CIPs remain constant in time because gravity is the only relevant interaction and acts equally on baryons and CDM.

The equal-time galaxy power spectrum $P_{gg}(k,z)$ is defined as $\langle{\delta}_g(\vk, z){\delta}_g(\vk', z)\rangle = 
P_{gg}(k,z) (2\pi)^3\delta_D(\vk + \vk')$, where $\delta_D$ is a Dirac delta and ${\delta}_g(\vk, z)$ is the Fourier 
transform of the galaxy overdensity. Using Eq.~(\ref{eq:biasexp}) this gives 
(dropping the dependence on redshift $z$ in the bias parameters to ease the notation)
\begin{equation}
\label{eq:Pgg}
\begin{split}
P_{gg}(k,z) &= b_1^2P_{mm}(k, z) + 2b_1b_\phi \fnl P_{m\phi}(k, z) + 2b_1b_\sigma P_{m\sigma}(k, z) \\
&\;\;\;\;+ b_\phi^2\fnl^2P_{\phi\phi}(k) + 2b_\phi b_\sigma \fnl P_{\phi\sigma}(k) 
+ b_\sigma^2P_{\sigma\sigma}(k) + P_{\eps\eps}(k)\,\,,
\end{split}
\end{equation}
where $P_{ab}$ denotes the cross-spectrum of the fields $a$ and $b$. 
In writing the above equation we have used the fact that the stochastic contribution $\eps$ 
does not correlate with any of the other fields.  
The density and primordial potential are related to the primordial scalar perturbation $\R(\vk)$, 
respectively, as $\delta_m(\vk, z) = (3/5)\mathcal{M}(k, z)\R(\vk)$ and $\phi(\vk) = (3/5) \R(\vk)$, 
where $\mathcal{M}(k, z)$ is defined as
\begin{equation}
\label{eq:M_def}
\mathcal{M}(k, z) = \frac{2}{3} \frac{k^2 T_m(k, z)}{\Omega_{m0}H_0^2},
\end{equation}
with $T_m$ being the total matter transfer function and $\Omega_{m0} = \Omega_{b0} + \Omega_{c0}$. 
The relation between $\sigma$ and $\R$ is not generically known a priori and depends on the mechanism that generates 
the CIP perturbation $\sigma$. Here, we take a model-independent approach and parametrize the amplitude of the CIPs 
at the level of their power spectrum. Specifically, we assume that the power spectrum of CIPs is close to scale-invariant with a tilt equal to that of the adiabatic perturbations; our analysis can be straightforwardly generalized to other spectral shapes. We thus write
\bq
\label{eq:Psigmasigma_1} P_{\sigma\sigma}(k) &=& \acip^2P_{\R\R}(k), \\
\label{eq:Psigmasigma_2} P_{\sigma\R}(k) &=& \xi \sqrt{P_{\sigma\sigma}(k)P_{\R\R}(k)} = \xi \acip P_{\R\R}(k)\,\,,
\eq
where $P_{\R\R}(k) = 2\pi^2 A_{\rm s}/k^3\left(k/k_{\rm p}\right)^{n_{\rm s} - 1}$ is the primordial adiabatic 
scalar perturbation power spectrum, $\acip$ is a parameter that describes the amplitude of the power of $\sigma$ 
and $\xi \in \left[-1, 1\right]$ quantifies the level of correlation between CIPs and adiabatic perturbations. 
With this parametrization, all of the contributions to Eq.~(\ref{eq:Pgg}) follow as
\bq
\label{eq:Pgg_contributions_1} P_{mm}(k,z) &=& \frac{9}{25} \mathcal{M}^2(k,z) P_{\R\R}(k)\,\,, \\
\label{eq:Pgg_contributions_2} P_{m\phi}(k,z) &=& \frac{9}{25} \mathcal{M}(k,z) P_{\R\R}(k)\,\,, \\
\label{eq:Pgg_contributions_3} P_{m\sigma}(k,z) &=& \frac{3}{5} \xi \acip \mathcal{M}(k,z) P_{\R\R}(k)\,\,, \\
\label{eq:Pgg_contributions_4} P_{\phi\phi}(k) &=& \frac{9}{25} P_{\R\R}(k)\,\,, \\
\label{eq:Pgg_contributions_5} P_{\phi\sigma}(k) &=& \frac{3}{5} \xi\acip P_{\R\R}(k)\,\,, \\
\label{eq:Pgg_contributions_6} P_{\sigma\sigma}(k) &=& \acip^2P_{\R\R}(k) \,\,, \\ 
\label{eq:Pgg_contributions_7} P_{\eps\eps}(k) &=& 1/\bar{n}_g\,\,,
\eq
where we have assumed Poisson statistics for the stochasticity $\epsilon$, with $\bar{n}_g$ being the mean galaxy number 
density observed within the surveyed volume. The contribution from each individual term in Eq.~(\ref{eq:Pgg}) 
is shown in Fig.~\ref{fig:model} (we evaluate all relevant spectra and transfer functions using the 
{\tt CAMB} code \cite{camb, 2011ascl.soft02026L}). Compared to the total matter contribution $P_{mm}$, 
the local PNG terms $P_{m\phi}$ and $P_{\phi\phi}$ introduce scale-dependent corrections $\propto \fnl k^{-2}$ 
and $\propto \fnl^2 k^{-4}$, respectively, which are the popular {\it scale-dependent bias} features of local 
PNG on the galaxy power spectrum \cite{dalal/etal:2008}. Equations~(\ref{eq:Pgg_contributions_1})--(\ref{eq:Pgg_contributions_7}) 
and Fig.~\ref{fig:model} reveal a point that is central to the discussion of this paper, 
which is that {\it CIPs contribute to the galaxy power spectrum with the same scale-dependence as local PNG.}
Concretely, for $\xi \neq 0$, CIPs contribute via $P_{m\sigma}$ with a term $\propto \xi\acip k^{-2}$ 
(cyan line in Fig.~\ref{fig:model}). If CIPs are assumed to be uncorrelated, $\xi = 0$, then the leading-order contribution 
is higher order in the amplitude of the CIP fluctuation: $P_{\sigma\sigma}/P_{mm} \propto \acip^2 k^{-4}$. 
These scale-dependent contributions are what can be exploited to constrain local PNG and primordial CIPs with the galaxy power spectrum. 

Before proceeding, we mention for completeness that photon-baryon interactions prior to the epoch of recombination will also naturally generate modulations of the relative abundance of baryons and CDM, $\delta_{bc}(\vx, z)$ \cite{barkana/loeb:11, 2016PhRvD..94f3508S, 2016ApJ...830...68A}. Here we ignore their contribution since the 
auto and cross-spectra of $\delta_{bc}$ with the other fields are completely negligible on the large scales (low $k$) 
at which we investigate the impact of $\fnl$ and $\acip$ (see e.g.~Fig.~3 of Ref.~\cite{2016PhRvD..94f3508S} 
or Fig.~6 of Ref.~\cite{2020JCAP...02..005B}). Besides, on BAO scales $k \sim 0.05 h/{\rm Mpc} - 0.1 h/{\rm Mpc}$, 
where $\delta_{bc}$ contributes the most, it does so by less than $1\%$ for a range of relevant redshifts and galaxy masses, 
as estimated in Ref.~\cite{2020JCAP...02..005B}. 
Neglecting the contribution from relative baryon-CDM density perturbations generated by photon-baryon 
interactions will therefore not have any impact on our conclusions. Similar arguments hold in the case of perturbations in the 
relative velocity between baryons and CDM \cite{tseliakhovich/hirata:2010, blazek/etal:15, 2016PhRvD..94f3508S}. 

The fact that the signatures of interest are only relevant on the largest observable scales justifies stopping at linear order in perturbations in Eq.~(\ref{eq:biasexp}). For simplicity, we also skip taking redshift space distortions (RSD) into account, but we note that this is not important to our main goal of demonstrating the comparable statistical errors that galaxy 
power spectra data can yield on $\fnl$ and $\acip$. Robust forecasts of the absolute values of $\sigma_{\fnl}$ and $\sigma_{\acip}$ should go beyond this simplifying assumption, but we do not perform these here. Further, on scales larger than the horizon $k \lesssim aH$, other {\it lightcone projection effects} from converting the observed galaxy distribution to that of their rest-frame also notoriously induce $\propto aH/k^2$ corrections to the galaxy power spectrum (see Ref.~\cite{2015CQGra..32d4001J} and Sec.~9.3 of Ref.~\cite{biasreview} for reviews). These effects contribute with the same scale-dependence as local PNG and correlated CIPs and, in principle, 
their amplitude can become important for $\fnl, \acip \lesssim 1$ (assuming order unity for all relevant bias parameters), which is close to the threshold sensitivity expected for future galaxy redshift surveys. Here, we skip including these terms for simplicity, and emphasize that more robust, survey-specific forecasts should examine in more detail their exact impact on the $\fnl$ and $\acip$ constraints. We stress, however, that the size of these corrections can be predicted given the evolution bias ${\rm d}\ln\bar{n}_g/{\rm d}\ln(1+z)$ and the luminosity function slope, hence, no degeneracies arise with $\fnl$ or $\acip$. 

%**********************
%**********************
\subsection{Multitracer power spectrum likelihood and covariance}\label{sec:model:cov}

When we forecast constraints on local PNG and primordial CIPs below, we work under the commonly 
adopted approximation of a Gaussian likelihood function
\bq\label{eq:likelihood}
\mathcal{L}(\fnl, \acip) \propto {\rm exp}\left[-\frac{1}{2}\Big({\bf M}(\fnl, \acip) 
- {\bf D}\Big)^t {\bf Cov}^{-1}\Big({\bf M}(\fnl, \acip) - {\bf D}\Big)\right]\,\,,
\eq
where ${\bf D}$ is the observed galaxy power spectrum data vector, ${\bf M}(\fnl, \acip)$ is 
the theoretical prediction and ${\bf Cov}$ is the covariance matrix of the data vector. 
We take as observables the bin-averaged power spectrum of the distribution of two galaxy samples $S_1$ and $S_2$ in real space, 
and their corresponding cross-spectrum. Our assumed data vector is given as
\bq\label{eq:datavector}
{\bf D}(k) = \Big\{\hat{P}_{gg}^{S_1S_1}, \hat{P}_{gg}^{S_1S_2}, \hat{P}_{gg}^{S_2S_2}\Big\}
\eq
with 
\bq\label{eq:angavePgg}
\hat{P}_{gg}^{S_1S_1}(k) &=& \frac{1}{V_SV_k} \int_{k} {\rm d}^3\vk'\ {\delta}_{g}^{S_1}(\vk'){\delta}_g^{S_1}(-\vk')\,\,, \\
\hat{P}_{gg}^{S_1S_2}(k) &=& \frac{1}{V_SV_k} \int_{k} {\rm d}^3\vk'\ {\delta}_{g}^{S_1}(\vk'){\delta}_g^{S_2}(-\vk')\,\,, \\
\hat{P}_{gg}^{S_2S_2}(k) &=& \frac{1}{V_SV_k} \int_{k} {\rm d}^3\vk'\ {\delta}_{g}^{S_2}(\vk'){\delta}_g^{S_2}(-\vk')\,\,,
\eq
where $V_S$ is the survey volume, $\int_k$ denotes averaging over a spherical shell in Fourier space centered at $k$ 
with width $\Delta k$ and volume $V_k = 4\pi k^2\Delta k + \O(\Delta k^3)$, and ${\delta}_g^{S_1, S_2}(\vk)$ 
are the Fourier modes estimated from the distribution of the two galaxy samples. In App.~\ref{app:estimator} 
we show that this estimator of the power spectrum is unbiased, i.e. 
\begin{equation}
\label{eq:unbiased}
\big<{\bf D}(k)\big> = \big\{P_{gg}^{S_1S_1}(k), P_{gg}^{S_1S_2}(k), P_{gg}^{S_2S_2}(k)\big\}\,\,.
\end{equation}
We also explicitly write the prediction for $P_{gg}^{S_1S_2}(k)$, 
which is a simple generalization of the expression for the auto-power spectrum in Eq.~(\ref{eq:Pgg}). 

The covariance matrix in Eq.~\eqref{eq:likelihood} is defined as ${\bf Cov} 
\equiv {\bf Cov}(k_1, k_2) = \big<{\bf D}(k_1){\bf D}(k_2)\big> - \big<{\bf D}(k_1)\big>\big<{\bf D}(k_2)\big>$. 
Ignoring the contribution from the connected four-point function piece, which is a good approximation 
on the scales we are interested in, the covariance is a $3\times3$ block-diagonal matrix given by 
(see App.~\ref{app:estimator} for the derivation)
\begin{equation}\label{eq:covariance}
{\bf Cov} = 2 \frac{(2\pi)^3\delta_{k_1k_2}}{V_SV_{k_1}}
\begin{pmatrix}
[P_{gg}^{S_1S_1}(k_1)]^2 & P_{gg}^{S_1S_2}(k_1)P_{gg}^{S_1S_1}(k_1) & [P_{gg}^{S_1S_2}(k_1)]^2 \\[1.5ex]
\cdots & \frac{\left[P_{gg}^{S_1S_1}(k_1)P_{gg}^{S_2S_2}(k_1) + [P_{gg}^{S_1S_2}(k_1)]^2\right]}{2} & P_{gg}^{S_2S_2}(k_1)P_{gg}^{S_1S_2}(k_1) \\[1.5ex]
\cdots & \cdots & [P_{gg}^{S_2S_2}(k_1)]^2
\end{pmatrix}
\,\,,
\end{equation}
where we skipped writing explicitly the entries marked with ``$\cdots$'' because the covariance matrix is symmetric.  

In all our forecast considerations below we consider $30$ $k$-bins equally spaced in log-scale between $\smash{k_F = 2\pi/V_S^{1/3}}$ 
and $k_{\rm max} = 0.1\ h/{\rm Mpc}$. We verified that our results do not depend sensitively on sensible changes to this setup. We also only vary $\fnl$ and $\acip$ in our forecast constraints and keep all remaining {cosmological and bias parameters fixed}: we place constraints on $\fnl-\acip$ space by evaluating the likelihood on a regular grid with wide linear flat priors on $\fnl$ and $\acip$. 

%**********************
%**********************
\section{Results}\label{sec:results}

In this section we present and discuss our numerical results. 
We start with a discussion on the impact of the bias parameters on the resulting $\fnl$ and $\acip$ constraints. 
We then examine how the constraints on $\fnl$ and $\acip$ depend on the correlation parameter $\xi$, 
as well as the improvements that result from incorporating CMB priors on $\fnl$. 
We also compare the constraining power of the galaxy power spectrum with existing constraints/forecasts on CIPs, and comment on the impact of a number of idealized aspects of our analysis. 

%**********************
%**********************
\subsection{The importance of galaxy bias for correlated CIPs, \texorpdfstring{$\xi = 1$}{\textbackslash xi = 1}}\label{sec:results:bias}

\begin{table*}%[!htbp]
\centering
\begin{tabular}{lcccc}
\toprule
%\rule{0pt}{1\normalbaselineskip}
 & Pair 1 & Pair 2 & Pair 3 & Pair 4 \\
\midrule
%\rule{0pt}{1\normalbaselineskip}
\\
\,\,$\big(z_{S_1},z_{S_2}\big)$						&	$\big(0.5, 0.5\big)$			&	$\big(1.0, 1.0\big)$			&	$\big(1.0, 1.0\big)$		&	$\big(2.0, 2.0\big)$ 
\\
\\
\,\,$\big(M_{\rm h}^{S_1}, M_{\rm h}^{S_2}\big)$	&	$\big(10^{12}, 10^{13}\big)$	&	$\big(5\times10^{12}, 5\times10^{13}\big)$	&	$ n.a. $						&	$ n.a. $
\\
\\
\,\,$\big(M_{*}^{S_1}, M_{*}^{S_2}\big)$			&	$ n.a. $						&	$ n.a. $						&	$\big(5\times10^{9}, 10^{11}\big)$	&	$\big(10^{9}, 5\times10^{10}\big)$
\\
\\
\,\,$\big(b_1^{S_1}, b_1^{S_2}\big)$				&	$\big( 0.99, 1.52 \big)$		&	$\big(1.81, 3.77\big)$			&	$\big(1.06, 1.81\big)$		&	$\big(1.43, 2.65\big)$ 
\\
\\
\,\,$\big(b_\phi^{S_1}, b_\phi^{S_2}\big)$			&	$\big( -0.04, 1.75\big)$		&	$\big(2.73, 9.33\big)$			&	$\big(0.19, 2.71\big)$		&	$\big(1.46, 5.58\big)$ 
\\
\\
\,\, $\big(b_\sigma^{S_1}, b_\sigma^{S_2}\big)$		&	$\big( -0.02, -0.31\big)$		&	$\big(-0.47, -1.42\big)$		&	$\big(0.32, 0.40\big)$		&	$\big(0.05, -0.12\big)$
\\[2ex]
\bottomrule
\end{tabular}
\caption{
  Specifications of the four pairs of galaxy samples used in this paper to discuss  the breaking of the degeneracy between $A$ and $\fnl$.  Pairs 1 and 2 (3 and 4) are selected by their total host halo mass $M_{\rm h}$ (stellar mass $M_*$),  i.e.~their bias parameters are determined by formulae given in terms of $M_{\rm h}$ ($M_*$);  see Sec.~\ref{sec:results:bias} for the bias parameter formulae as a function of $M_{\rm h}$ and $M_*$.  The masses are quoted in units of $M_\odot/h$. In all pairs, we crudely account for the sparsity of higher mass galaxies by assuming sample $S_2$ to contain $5\%$ of the total number of observed galaxies $\bar{n}_g$, i.e., $\bar{n}_g^{S_2} = 0.05\bar{n}_g$, $\bar{n}_g^{S_1} = \bar{n}_g - \bar{n}_g^{S_2}$.}
\label{table:pairs}
\end{table*}

The contributions from $\fnl$ and $\acip$ to the galaxy power spectrum are largely degenerate with 
each other for fully correlated CIPs ($\xi = 1$), making it nontrivial to obtain tight constraints on either one of the parameters. 
To leading order in $\fnl$ and $\acip$ the degeneracy is actually perfect. 
This can be seen by inspecting the $P_{m\phi}$ and $P_{m\sigma}$ terms in Eq.~(\ref{eq:Pgg}), 
whose joint contribution cancels exactly if
\bq\label{eq:degdir}
\acip = -\frac{3}{5} \frac{b_\phi}{b_\sigma} \fnl\,\,.
\eq
The degeneracy is broken by the second-order terms in $\fnl$ and $\acip$ ($P_{\phi\phi}, P_{\phi\sigma}, P_{\sigma\sigma}$), 
which have a different scale dependence, but this happens efficiently only when $\fnl$ and $\acip$ are large. 

One way to break this degeneracy for correlated CIPs is by carrying out joint analyses of the power spectrum of 
at least two galaxy samples $S_1$ and $S_2$ with different bias values. 
This is a technique referred to as the galaxy multitracer technique \cite{2009JCAP...10..007M, 2009PhRvL.102b1302S}, 
which comes also with the added benefit that if the two galaxy samples cover the same volume of the Universe, then sample variance errors can cancel to a large extent. 
With two galaxy samples at hand, it becomes possible to probe the $\fnl-\acip$ space along two different directions:
\begin{equation}
\label{eq:degdir_2}
\text{$\acip = -\frac{3}{5} \frac{b_\phi^{S_1}}{b_\sigma^{S_1}} \fnl$ \quad and \quad 
$\acip = -\frac{3}{5} \frac{b_\phi^{S_2}}{b_\sigma^{S_2}} \fnl\,\,;$}
\end{equation}
the superscripts $^{S_1}$ and $^{S_2}$ label the bias values of the two galaxy samples. 
The degeneracy can therefore be broken efficiently if (i) the ratios 
$b_\phi^{S_1}/b_\sigma^{S_1}$, $b_\phi^{S_2}/b_\sigma^{S_2}$ are sufficiently different; and (ii) all four bias parameters 
remain sizable to keep the galaxy power spectrum sensitive to $\fnl$ and $\acip$. 

To organize the discussion about the importance of galaxy bias we have chosen four example pairs of galaxy samples 
with the specifications listed in Table \ref{table:pairs}. 
The bias values for each of the galaxy samples is determined as follows. 
We say that the galaxy samples in pairs $1$ and $2$ are selected by their total host halo mass $M_{\rm h}$, 
i.e., we evaluate their bias parameters using formulae given in terms of $M_{\rm h}$. For $b_1(M_{\rm h})$ 
we use the fitting function of Ref.~\cite{2010ApJ...724..878T} with the parameters listed in their Table 2. The galaxy bias parameter associated with local PNG can be evaluated by computing the response of galaxy number counts to long-wavelength spatial modulations of the amplitude of the primordial adiabatic scalar power spectrum (which is what characterizes primordial non-Gaussianity of the local type \cite{dalal/etal:2008, slosar/etal:2008}). Using the separate universe ansatz, we can write
\bq\label{eq:bphi_tinker}
b_{\phi}(M_{\rm h}) = \frac{4}{\Delta A_{{\rm s},L}}\left[\frac{n^{{\rm SepUni, \fnl}}(z,M_{\rm h})}{n^{{\rm Fiducial}}(z,M_{\rm h})} - 1\right]\,\,,
\eq
where the separate universe and the fiducial cosmologies have the same cosmological parameters, except $A_{\rm s}$ which differs by $\Delta A_{{\rm s},L}$. Under the assumption of universality of the halo mass function \cite{slosar/etal:2008, 
ferraro/etal:2012, PBSpaper, scoccimarro/etal:2012, 2017MNRAS.468.3277B} it follows from the above equation that $b_{\phi}(M_{\rm h}) = 2\delta_{c}\big(b_1(M_{\rm h}) - 1\big)$, where $\delta_{c} = 1.686$ is the linearly-extrapolated to $z=0$ threshold overdensity for the collapse of a spherical perturbation in $\Lambda{\rm CDM}$; this is the formula we adopt in our results. Finally, for $b_{\sigma}$ we follow Ref.~\cite{2020JCAP...02..005B} and evaluate it using also the separate universe ansatz as
\bq\label{eq:bsigma_tinker}
b_{\sigma}(M_{\rm h}) = \frac{1}{\sigma_{L}}\left[\frac{n^{{\rm SepUni,CIP}}(z,M_{\rm h})}{n^{{\rm Fiducial}}(z,M_{\rm h})} - 1\right]\,\,,
\eq
where $\sigma_{L}$ is the amplitude of an infinite-wavelength primordial CIP 
and the separate universe and fiducial cosmologies have the same parameters, except $\smash{\Omega_{b0}^{\rm SepUni,CIP} 
= \Omega_{b0}^{\rm Fiducial}\left[1 + \sigma_{L}\right]}$ and $\smash{\Omega_{c0}^{\rm SepUni,CIP} 
= \Omega_{c0}^{\rm Fiducial}\left[1 - f_b \sigma_{L}\right]}$, 
with $\smash{f_b = \Omega_{b0}^{\rm Fiducial}/\Omega_{c0}^{\rm Fiducial}}$.  Equation (\ref{eq:bsigma_tinker}) was shown in Ref.~\cite{2020JCAP...02..005B} to successfully reproduce 
the $b_\sigma$ values measured for halos in gravity-only simulations (see their Fig.~1).\footnote{Note, in Ref.~\cite{2020JCAP...02..005B}, their galaxy bias parameter is denoted by $b_{\delta}^{bc}$ and it is related to the CIP bias parameter here as $b_\sigma = (1 + f_b) b_{\delta}^{bc}$.} We evaluate 
the halo abundances $n^{{\rm Fiducial}}(z,M_{\rm h})$, $n^{{\rm SepUni}}(z,M_{\rm h})$ in 
the fiducial and separate universe cosmologies, respectively, using the fitting formulae of 
Ref.~\cite{2008ApJ...688..709T} for the spherical-overdensity $\Delta = 200$ definition (see their Table 2). 

{On the other hand, galaxy pairs $3$ and $4$ in Table \ref{table:pairs} have bias values representative of galaxy samples selected by stellar mass; this is more realistic observationally as galaxy stellar mass determinations are more robust than estimates of the corresponding total host halo mass. Our bias determination in this case is, however, necessarily more approximate due to the fewer number of galaxy bias estimates as a function of stellar mass $M_*$ in the literature. A recent example, however, }is the work of Ref.~\cite{2020JCAP...02..005B}, who measured $b_{\sigma}$ as a function of $M_{*}$ for galaxies simulated with the IllustrisTNG model 
\cite{2017MNRAS.465.3291W, Pillepich:2017jle} (see also Refs.~\cite{2010ApJ...710..903M, 2018MNRAS.475..676S} for investigations of the dependence of $b_1$ on quantities beyond just host halo mass). There it was found that simple analytical modeling of 
the stellar-to-halo mass relation can reproduce well the results measured from the simulations. 
Concretely, we evaluate $b_{\sigma}(M_*)$ using the following generalization of Eq.~(\ref{eq:bsigma_tinker}),
\bq\label{eq:bsigma_tinker_stemass}
b_{\sigma}(M_{*}) = \frac{1}{\sigma_{L}}\left[\frac{n^{{\rm SepUni,CIP}}(z,M_{\rm h}
[(1-\eps_*\sigma_L)M_*])}{n^{{\rm Fiducial}}(z,M_{\rm h}[M_*])} - 1\right]\,\,,
\eq
where $M_{\rm h}[M_*]$ represents a fit to the mean relation between total host halo mass 
and stellar mass found in the simulations of Ref.~\cite{2020JCAP...02..005B} and $\eps_* = 0.75$ is a fitting parameter 
that describes how much more mass in stars there is at fixed halo mass given a positive linear CIP. 
On the other hand, given the lack of precise priors on the relation between $b_1$ and $b_\phi$ as a function of stellar mass, 
here we evaluate them using the same formulae as for total host halo mass by simply replacing $M_h \rightarrow M_{\rm h}[M_*]$. 
Physically, this amounts to assuming that the stellar-to-halo mass relation is unaffected by the presence of a long-wavelength 
total matter perturbation $\delta_m$ and local PNG. 
We proceed with this assumption, but keep in mind that it should 
eventually be tested with hydrodynamical simulations of galaxy formation \cite{barreirainprep}. 

\begin{figure}[t!]
		\centering
		\includegraphics[width=\textwidth]{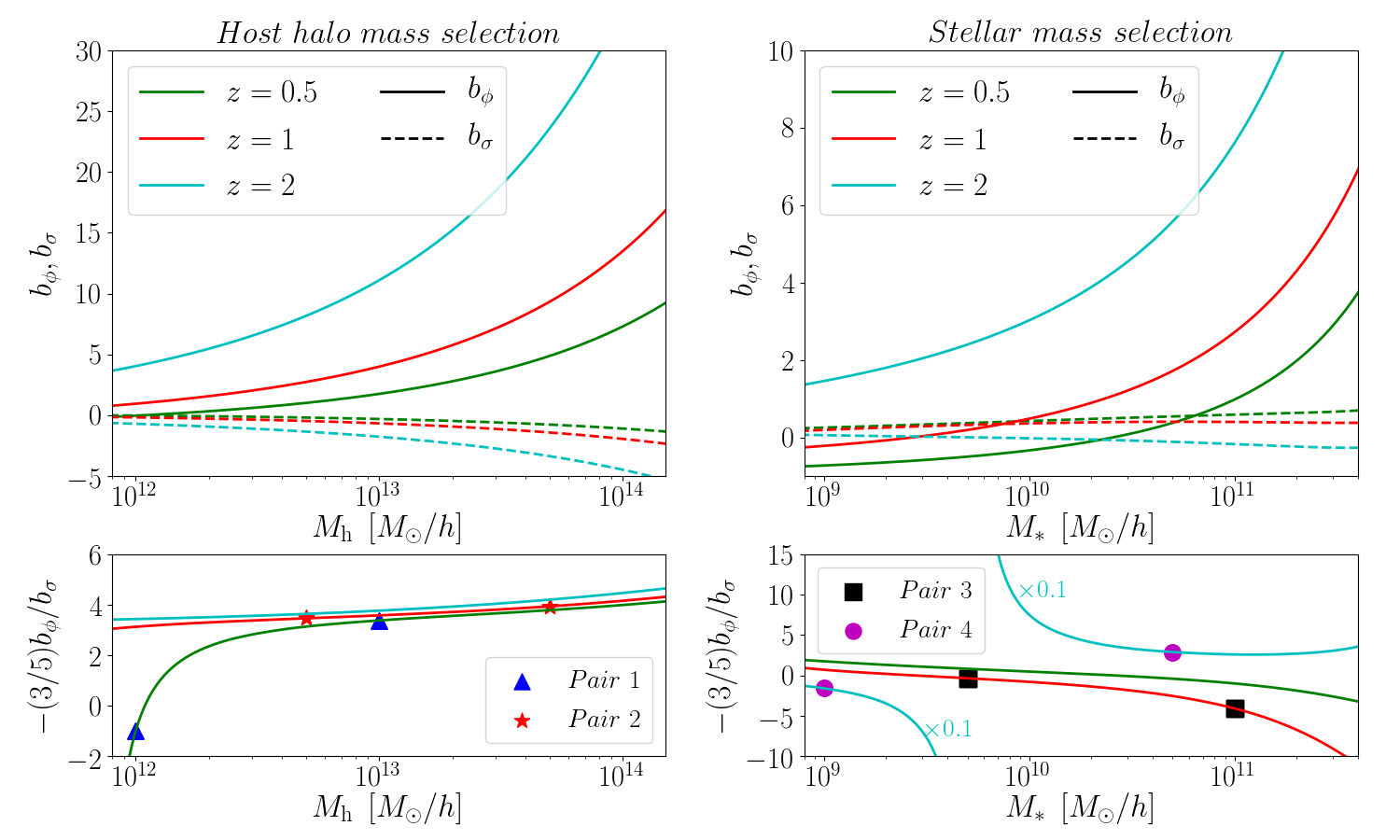}
		\caption{ 
		Galaxy bias parameters $b_\phi$ (solid) and $b_\sigma$ (dashed) as a function of total host halo mass (left) 
		and stellar mass (right) and for different redshifts, as labeled. 
		The upper panels show the bias values, while the lower panels show the degeneracy directions in $\fnl-\acip$ space, 
		$-(3/5)b_\phi/b_\sigma$ (cf.~Eq.~(\ref{eq:degdir})). 
		The symbols in the lower panels mark the mass and degeneracy direction values of the galaxy samples of the pairs listed in 
		Table \ref{table:pairs}. To improve visualization, the $z=2$ curve in the lower right panel is scaled by a factor of $0.1$; this curve also diverges at $M_* \approx 6\times 10^{9}M_{\odot/h}$ as $b_\sigma = 0$ there.}
\label{fig:bias_intuition}
\end{figure}

\begin{figure}[t!]
		\centering
		\includegraphics[width=\textwidth]{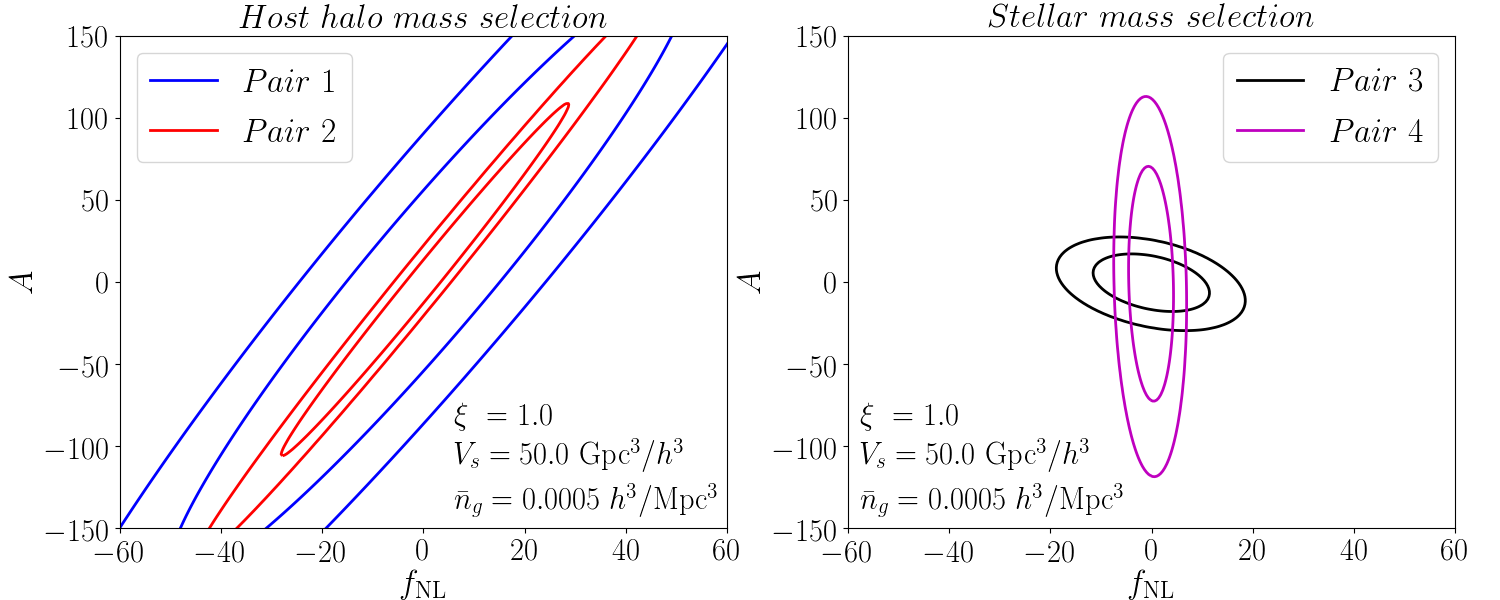}
		\caption{Forecasted constraints on $\fnl$ and $\acip$ for the four galaxy sample pairs listed in Table \ref{table:pairs}, as labeled; 
		the contours mark $1\sigma$ and $2\sigma$ confidence regions. 
		The result corresponds to fully correlated CIPs, $\xi = 1$, and for a survey volume $V_S = 50\ {\rm Gpc}^3/h^3$. 
		For all four pairs, the B sample is assumed to contain $5\%$ of the total number of galaxies, $\bar{n}_g^{S_2} = 0.05\bar{n}_g, \bar{n}_g^{S_1} = \bar{n}_g - \bar{n}_g^{S_2}$.}
\label{fig:contours_intuition}
\end{figure}

The host halo and stellar mass dependence of $b_\phi$ and $b_{\sigma}$ is shown in the upper panels of Fig.~\ref{fig:bias_intuition}, 
as labeled. The markers in the lower panels indicate the mass and degeneracy direction, $-(3/5)b_\phi/b_\sigma$, 
of the galaxy samples of the four pairs listed in Table \ref{table:pairs}. 
The corresponding constraints on $\fnl-\acip$ space obtained with each pair are depicted in Fig.~\ref{fig:contours_intuition} 
for $\xi = 1$, $V_S = 50 {\rm Gpc}^3/h^3$ and $\bar{n}_g^{S_2} = 0.05\bar{n}_g, \bar{n}_g^{S_1} = \bar{n}_g - \bar{n}_g^{S_2}$, which shows that the stellar mass selected pairs yield tighter constraints compared to the halo mass selected ones.
This can be explained by noting that the degeneracy direction $-(3/5)b_\phi/b_\sigma$ does not depend strongly on 
both redshift and total host halo mass for sizable values of $b_\phi$ and $b_\sigma$; 
for $z = 1$ and $z=2$, $-(3/5)b_\phi/b_\sigma$ varies from $\approx 3$ to $\approx 4$ in between $M_{\rm h} = 10^{12}\ M_{\odot}/h$ 
and $M_{\rm h} = 10^{14}\ M_{\odot}/h$. In other words, regardless of the redshift and typical host halo mass of the galaxies chosen, 
the two degeneracy directions in Eq.~(\ref{eq:degdir_2}) are approximately the same, 
and the resulting constraints on $\fnl$ and $\acip$ are weaker because the degeneracy is only weakly broken. 
The case of pair $1$ at $z=0.5$ is an interesting one because, although each of its samples probes very distinct degeneracy 
directions (cf.~blue triangles in the lower left panel of Fig.~\ref{fig:bias_intuition}), 
both $b_\phi$ and $b_\sigma$ are small for $M_{\rm h} = 10^{12}\ M_{\odot}/h$. 
In this case, the degeneracy is formally broken, but at the price of having one of the galaxy samples very weakly {\it responsive} 
to $\fnl$ and $\acip$ and hence very poorly constraining; one is effectively left with a strong degeneracy that is similar to what one would have obtained with a single galaxy sample.

On the other hand, for the case of stellar mass selection, it is possible to construct pairs whose samples have sufficiently different degeneracy directions 
$-(3/5)b_\phi/b_\sigma$ for sizable $b_\phi$ and $b_\sigma$ values.  The case of pair 3 at $z = 1$ is illustrative of this, and hence it can break the degeneracy efficiently and yield tight confidence levels in $\fnl-\acip$ space (black contours in the right panel of Fig.~\ref{fig:contours_intuition}). The example of the stellar mass-selected pair 4 at $z = 2$ is also an interesting one. For this case, the two degeneracy directions are sufficiently different, but the value of $b_{\sigma}$ for $M_* = 10^9 M_{\odot}/h$ at $z=2$ is small, making the corresponding sample weakly {\it responsive} to $\acip$. The result is a weak degeneracy, but with degraded constraints on $\acip$ (worse by a factor of $\approx5$, compared to pair 3), as shown by the magenta contours on the right panel of Fig.~\ref{fig:contours_intuition}.

The galaxy bias parameters in general depend on the complicated and uncertain processes associated with galaxy formation and evolution, and cosmological inference analyses using the shape of the power spectrum benefit strongly from theoretical priors on their amplitude and time evolution. The discussion here shows that the importance of such priors is made even greater for the case of joint constraints on PNG and correlated CIPs because of their degeneracy; an improved knowledge of galaxy bias can in fact be used to determine which pairs of galaxy samples are able to return optimal constraints on $\fnl-\acip$ space. We note, for example, that in all our example galaxy sample pairs in Table \ref{table:pairs}, both samples are taken at the same redshift, but the $\fnl - \acip$ degeneracy can also be broken by choosing samples at different redshifts. In this case, however, one would not take advantage of the sample variance cancellation as the two galaxy samples would cover different regions of the universe.

%**********************
%**********************
\subsection{The impact of CIP correlation \texorpdfstring{$\xi$}{\textbackslash xi}}\label{sec:results:xi}

\begin{figure}[t!]
		\centering
		\includegraphics[width=\textwidth]{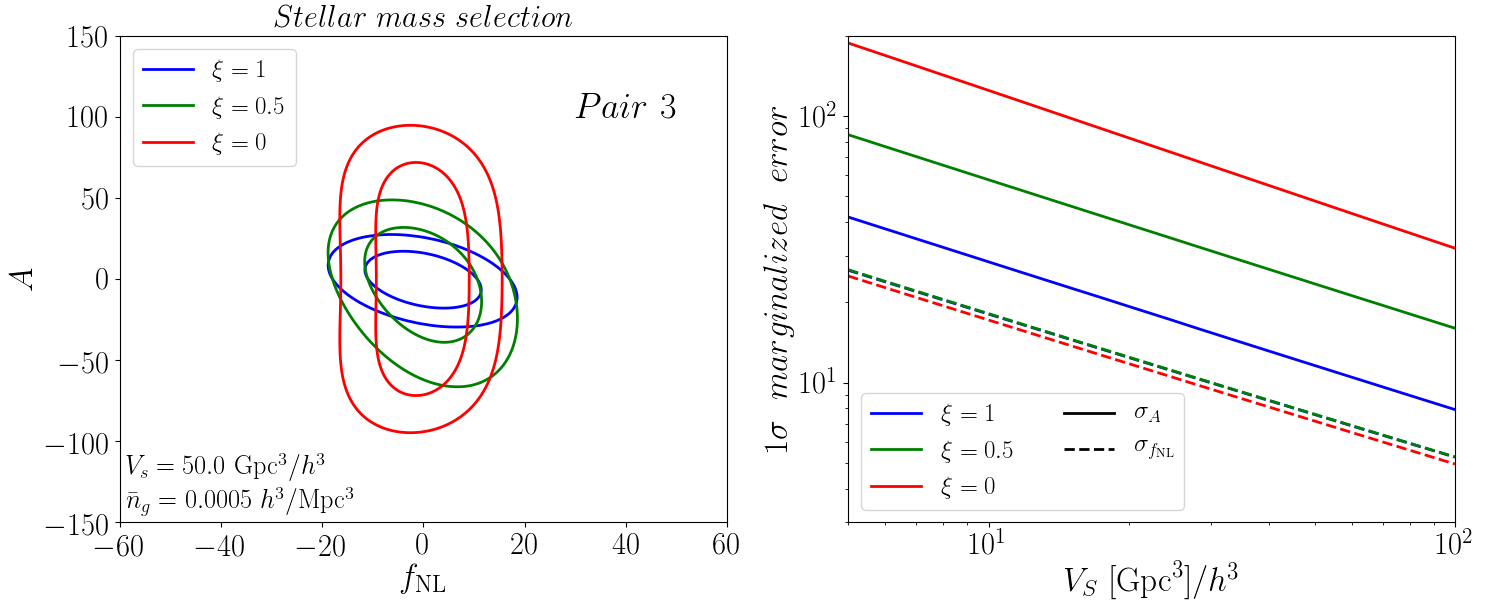}
		\caption{Impact of the CIP correlation parameter $\xi$ on $\fnl$ and $\acip$ constraints. 
		The left panel shows the constraints on $\fnl - \acip$ space for $\xi = 1$, $\xi = 0.5$ and $\xi = 0$, as labeled; 
		the result is for the stellar mass-selected galaxy sample pair $3$ in Table \ref{table:pairs}, $V_S = 50 {\rm Gpc}^3/h^3$, 
		$\bar{n}_g^{S_2} = 0.05\bar{n}_g, \bar{n}_g^{S_1} = \bar{n}_g - \bar{n}_g^{S_2}$ and the contours mark $1\sigma$ and $2\sigma$ confidence regions. 
		The right panel shows the corresponding marginalized $1\sigma$ constraints on $\fnl$ (dashed) 
		and $\acip$ (solid) as a function of survey volume $V_S$; the blue and green dashed curves are nearly overlapping.}
\label{fig:xipair3}
\end{figure}

\begin{figure}[t!]
		\centering
		\includegraphics[width=\textwidth]{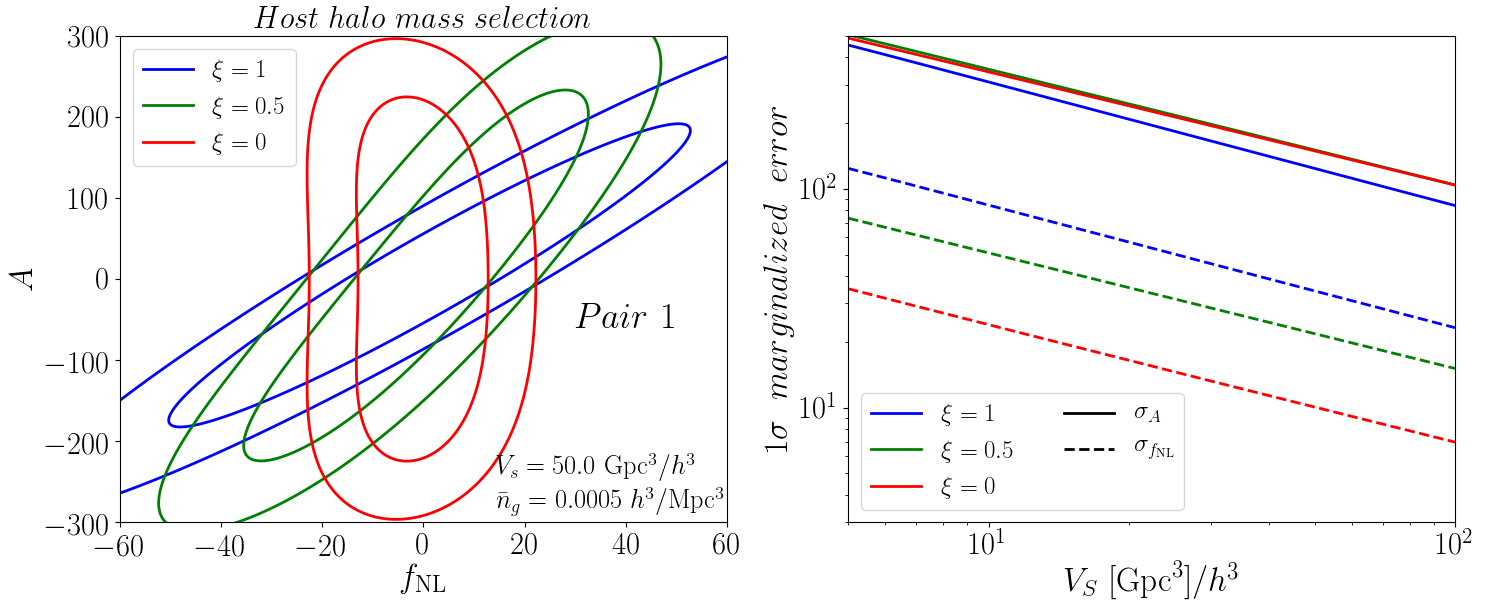}
		\caption{Same as Fig.~\ref{fig:xipair3}, but for the host halo mass-selected galaxy sample pair $1$ in Table \ref{table:pairs}.} 
\label{fig:xipair1}
\end{figure}

Figure~\ref{fig:xipair3} illustrates the impact of different values of the CIP correlation coefficient 
$\xi$ on the $\fnl$ and $\acip$ constraints using galaxy pair $3$ in Table \ref{table:pairs}. 
As $\xi \rightarrow 0$, the contribution from the $P_{m\sigma}$ term in Eq.~(\ref{eq:Pgg}) becomes smaller and 
thus the degeneracy between $\fnl$ and $\acip$ weaker. Concretely, in the limit $\xi=0$, 
the constraints on $\fnl$ come mostly from the $P_{m\phi}$ term in Eq.~(\ref{eq:Pgg}), 
while those on $\acip$ come solely from the $P_{\sigma\sigma}$ term; these two terms have a distinct scale dependence 
(cf.~Fig.~\ref{fig:model}), which results in effectively no degeneracy between $\fnl$ and $\acip$, 
as illustrated by the red contours in Fig.~\ref{fig:xipair3}. For the specific case of galaxy sample pair 3, the constraints on $\fnl$ barely change and those on $\acip$ worsen by about a factor of 4 from $\xi = 1$ to $\xi = 0$.

If for $\xi = 0$ the degeneracy between $\fnl$ and $\acip$ is no longer present, it follows that galaxy bias 
does not have as critical an importance in the resulting constraints, compared to the $\xi = 1$ case discussed 
in the previous subsection. This is illustrated in Fig.~\ref{fig:xipair1}, which shows the same as Fig.~\ref{fig:xipair3},
but for galaxy sample pair $1$ in Table \ref{table:pairs}. When $\xi = 1$ (blue), the strong degeneracy that we discussed in Sec.~\ref{sec:results:bias} results in poor constraints on both $\fnl$ and $\acip$. On the other hand, as $\xi \to 0$, the constraints on $\acip$ worsen just slightly, but those on $\fnl$ improve significantly (approximately a factor of $3.5$ from $\xi = 1$ to $\xi = 0$) because of the weaker degeneracy between the two physical effects.

%**********************
%**********************
\subsection{The impact of CMB priors on \texorpdfstring{$\fnl$}{f\_\{NL\}}}\label{sec:results:cmbprior}

\begin{figure}[t!]
		\centering
		\includegraphics[width=\textwidth]{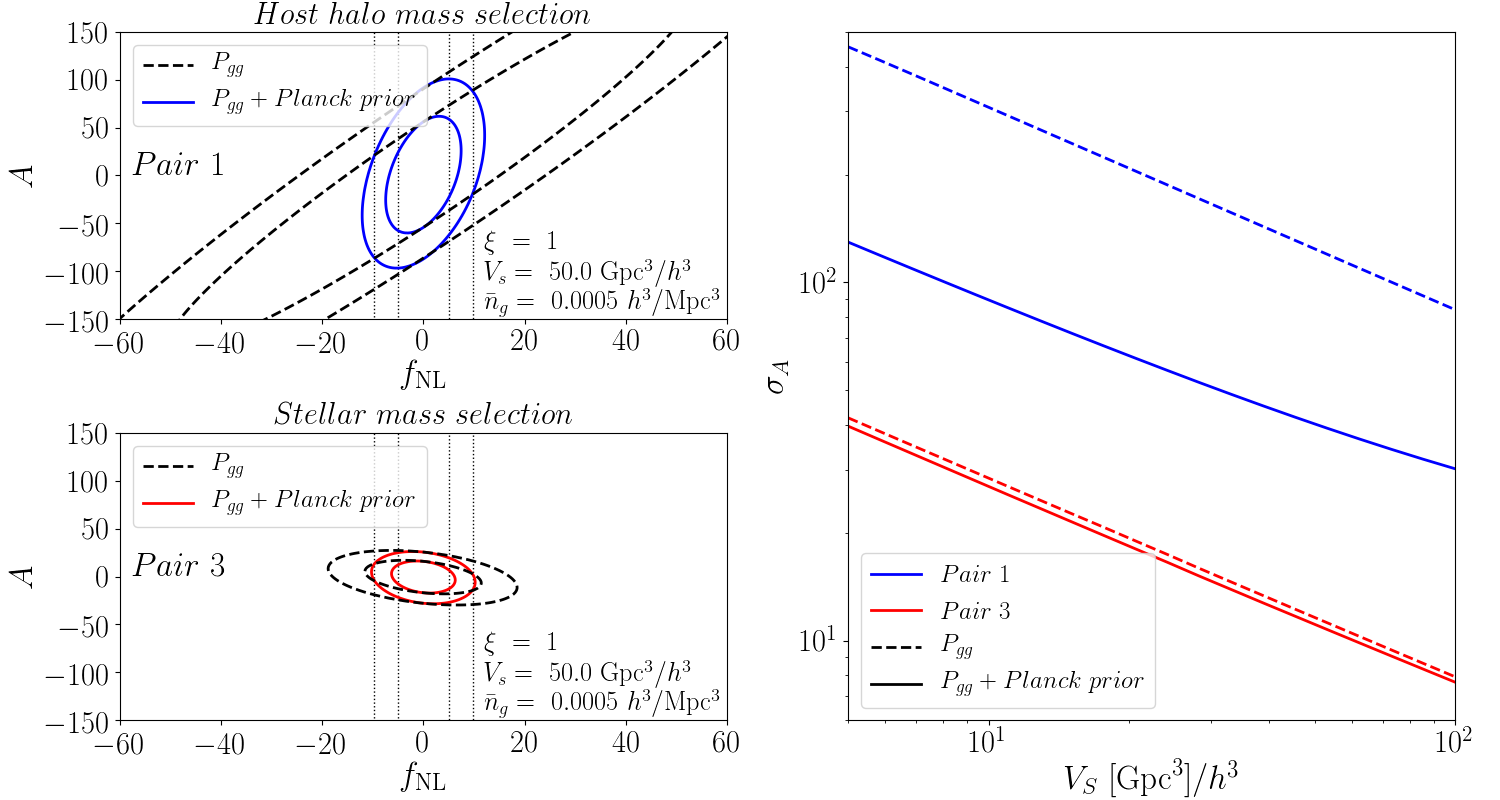}
		\caption{Impact of CMB priors from Planck on $\fnl$ and $\acip$ constraints (for $\xi = 1$) using the galaxy power spectrum $P_{gg}$. 
		The left panels show the constraints from galaxy sample pairs $1$ and $3$ on $\fnl-\acip$ space without (dashed black) and 
		with (solid) a Planck prior on $\fnl$ taken into account, as labeled; 
		the contours without the Planck prior are the corresponding ones in Fig.~\ref{fig:contours_intuition}. 
		The $\fnl$ prior from the CMB is taken here to be a Gaussian with mean $0$ and standard deviation $\sigma_{\fnl}^{\rm CMB} = 5$ 
		(the vertical dotted lines mark these $1\sigma$ and $2\sigma$ intervals).
		The right panel shows the marginalized $1\sigma$ constraints on $\acip$ as a function of survey volume $V_S$.}
\label{fig:fnlprior}
\end{figure}

The latest results from the Planck satellite provide currently the strongest bounds on local PNG and 
they constrain $\fnl = -0.9 \pm 5.1\ (1\sigma)$ \cite{2019arXiv190505697P}. The degenerate effects of $\fnl$ and $\acip$ 
on the galaxy power spectrum discussed above for correlated CIPs in Sec.~\ref{sec:results:bias} raise the interesting question of how much CIP constraints improve when including prior CMB information on $\fnl$. 
This is illustrated in Fig.~\ref{fig:fnlprior}, which shows the impact of a Gaussian prior on $\fnl$ with mean $0$ 
and standard deviation $\sigma_{\fnl}^{\rm CMB} = 5$ on the constraints. 
As it would be expected, adding a prior on $\fnl$ generically results in improved constraints, 
but the size of the improvement depends on the specific galaxy sample pairs at hand. 
Concretely, as we have seen in Sec.~\ref{sec:results:bias}, pair $1$ alone cannot constrain $\acip$ nor $\fnl$ efficiently 
for $\xi = 1$ because of their strong degeneracy (dashed contours in the upper left panel), 
but relatively tight constraints become possible once the CMB prior on $\fnl$ is added (solid contours in the upper left panel and solid blue vs.~dashed blue curves in the right panel of Fig.~\ref{fig:fnlprior}).\footnote{The Planck constraints on $\fnl$ were obtained under the assumption of $\acip = 0$. One might wonder whether the use of this prior is self-consistent, in particular in the case $\xi=1$. We argue however that correlated CIPs do not affect the Planck constraint on $\fnl$. The latter is dominated by the squeezed CMB bispectrum, whose leading contributions from both $\fnl$ and $\acip$ can be written as $B(\ell_L, \ell_S, \ell_S) = R(\ell_S) C({\ell_L}) C({\ell_S})$, where $\ell_L \ll \ell_S$, $C(\ell)$ is the angular power spectrum and $R(\ell_S)$ a response function. For $\fnl$, $R(\ell_S) \propto {\rm d}\ln C({\ell_S})/{\rm d}\ln {A}_{\rm s}$, while for a correlated CIP mode, $R(\ell_S) \propto {\rm d}\ln C({\ell_S})/{\rm d}\ln (\Omega_b h^2)$. The $\ell_S$-dependence of the $\fnl$ and $\acip$ contributions are therefore very different, thereby making CMB constraints on $\fnl$ only weakly dependent on $\acip$, even if CIP and adiabatic modes are fully correlated.} This breaking of the $\fnl-\acip$ degeneracy by the CMB prior also implies that the constraints become less critically dependent on galaxy bias to break the degeneracy. Notice that we do not include CMB prior information on $\acip$, since CMB constraints on $\acip$ are so poor in comparison that the prior would not make a difference in forecasted constraints. 

On the other hand, the CMB prior has a reduced importance for a stellar mass selected galaxy sample pair like pair 3 in Table \ref{table:pairs}, which is capable of tighter constraints on both $\fnl$ and $\acip$ already (lower left panel of Fig.~\ref{fig:fnlprior}). In fact, because of the rather weak degeneracy in $\fnl-\acip$ space for this pair, the constraints on $\acip$ are only slightly affected by the shrinkage of the contours along the $\fnl$ direction by the CMB prior, for all of the $V_S$ values probed.

%**********************
%**********************
\subsection{Comparison to existing CIP constraints and forecasts}
\label{sec:results:comparison}

In this subsection, we compare the constraining power of the 
galaxy power spectrum with existing constraints and other forecasts on CIPs. 
The latter are often quoted in terms of their root-mean-square CIP amplitude over some scale $R$, $\Delta^2_{\rm rms}(R)$, defined as 
\bq\label{eq:rms}
\Delta^2_{\rm rms}(R) = \frac{1}{2\pi^2}\int{\rm d}k\, k^2 \left(\frac{3j_1(kR)}{kR}\right)^2 P_{\sigma\sigma}(k)\,\,.
\eq
The tightest constraints to date on CIPs come from CMB data and they are obtained 
for the case of uncorrelated CIPs with a scale-invariant power spectrum, $\smash{P_{\sigma\sigma}(k) = A^{\rm S.I.}/k^3}$. 
The scale-invariant parametrization of the CIP power spectrum is approximately the same as our parametrization of CIPs 
in Eq.~(\ref{eq:Psigmasigma_1}), as the power spectrum of the adiabatic scalar perturbations is close to scale-invariant anyway 
($n_{\rm s} \approx 1$)
\bq
\label{eq:Psigmasigma_1_approx}
P_{\sigma\sigma}(k) = \acip^2 P_{\R\R}(k) = 
\frac{2\pi^2 \acip^2 A_{\rm s}}{k^3}\left(\frac{k}{k_{\rm p}}\right)^{n_{\rm s} - 1} 
\approx \frac{2\pi^2 \acip^2 A_{\rm s}}{k^3}\,\,,
\eq 
from which we obtain $A^{\rm S.I.}\approx 2\pi^2 \acip^2 A_{\rm s}\approx 4\times 10^{-8} \acip^2$. Some examples of bounds include $\Delta^2_{\rm rms}(R_{\rm CMB}) < 0.0043\ (2\sigma)$ \cite{2017PhRvD..96h3508S} 
and $\Delta^2_{\rm rms}(R_{\rm CMB}) < 0.005\ (1\sigma)$ \cite{2016PhRvD..93d3008M}, 
as well as the analysis of Ref.~\cite{2017JCAP...04..014V} 
which finds hints for non-zero CIPs, $\Delta^2_{\rm rms}(R_{\rm CMB}) = 0.0069^{+0.0030}_{-0.0031}\ (1\sigma)$. 
The analysis of the Planck collaboration \cite{2018arXiv180706211P} also reports a non-zero detection at the $1 \sigma$ level: 
$\Delta^2_{\rm rms}(R_{\rm CMB}) = 0.0037^{+0.0016}_{-0.0021}\ (1\sigma)$ (cf.~{\tt TT, TE, EE + lowE + lensing (conserv.)} 
data combination in their Table 15). For all these constraints, $R_{\rm CMB}$ is of order the size of the sound horizon at recombination, for which $\Delta^2_{\rm rms}(R_{\rm CMB})
\approx 10^{-8}\acip^2$ from Eq.~(\ref{eq:rms}) with $k_{\rm min} \sim 10\ {\rm Gpc}^{-1}$ as the lower integration limit (see also Sec.~IV of Ref.~\cite{2017PhRvD..96h3508S} for a summary of past constraints on CIPs).\footnote{Another competitive bound is one that can be achieved by looking for spatial modulations of the 
baryon fraction in galaxy clusters \cite{2010ApJ...716..907H}. 
This constrains $\Delta^2_{\rm rms}(R = 10\ {\rm Mpc}/h) < 0.006\ (1\sigma)$, which corresponds 
to $\acip \lesssim 650\ (1\sigma)$ \cite{grin/dore/kamionkowski}.}

For comparison purposes below, we take the $2\sigma$ bound from Ref.~\cite{2017PhRvD..96h3508S} divided by $2$ 
as representative of the current typical $1\sigma$ upper limit on CIPs, which implies
\bq\label{eq:cmbconstraints}
{\acip} \lesssim 450 \qquad {\rm (current\ CMB\ constraints)}\,\,.
\eq
This is a number that can be contrasted with the constraints we have shown above in Sec.~\ref{sec:results:xi} for $\xi = 0$ (but keeping in mind the simplified nature of our analysis). 
Concretely, one can read from the right panel of Fig.~\ref{fig:xipair3} that galaxy sample pairs like pair $3$ 
in Table \ref{table:pairs} are able to reach bounds of ${\acip} \lesssim 50$ for a survey size of $V_S = 50\ {\rm Gpc}^3/h^3$.
As discussed above, the constraining power of the galaxy power spectrum depends on the values of galaxy bias;
for example, for pair $1$ in Table \ref{table:pairs} the same bound would be $\acip \lesssim 150$ (right panel of Fig.~\ref{fig:xipair1}).

The constraining power of the galaxy power spectrum gets even stronger for the case of correlated CIPs $\xi = 1$, 
although for this case a more careful treatment of the galaxy selection variables and the resulting bias parameters is beneficial 
to break the stronger degeneracies that can arise between $\fnl$ and $\acip$ (cf.~Fig.~\ref{fig:contours_intuition}). 
Of the four pairs of galaxy samples in Table \ref{table:pairs}, pair $1$ is the one 
with the strongest degeneracy for $\xi = 1$, but even for this case bounds of ${\acip} \lesssim 40$
are possible if a survey volume $V_S = 50\ {\rm Gpc}^3/h^3$ is combined with a prior on $\fnl$ from Planck 
(cf.~blue curve on the right panel of Fig.~\ref{fig:fnlprior}). 
Galaxy sample pair $3$ is representative of stellar mass-selected cases that efficiently break the degeneracy between $\fnl$ and $\acip$ (cf.~Fig.~\ref{fig:contours_intuition}) and yields the tightest constraints of our analysis; for volumes $V_S \gtrsim 50\ {\rm Gpc}^3/h^3$, galaxy sample pair 3 is able to constrain ${\acip} \lesssim 10$.

The galaxy power spectrum is therefore also a strong probe of correlated CIPs ($\xi = 1$), but other probes, 
including the CMB, are expected to return comparably tight constraints. 
For instance, Ref.~\cite{2015PhRvD..92f3018H} studies the effect of correlated CIPs on the CMB and 
forecasts that constraints of order {$\sigma_{\acip} \approx 20$ could be achieved with Planck data already. Reference \cite{2018PhRvD..97b3513H} further shows that cosmic-variance limited temperature and polarization data could probe $\sigma_A \sim 6$.}  More recently, Ref.~\cite{2019arXiv190808953H} studied the impact of correlated CIPs on the cross-correlation 
of the galaxy distribution and kSZ effect data that arises from the same $\propto b_\sigma\sigma(\vx)$ contribution 
to the galaxy overdensity $\delta_g(\vx, z)$ that we studied here. 
The authors of Ref.~\cite{2019arXiv190808953H} find that the combination of galaxy and kSZ data can 
also constrain CIP amplitudes comparable to those of adiabatic perturbations; one of their best-case scenarios 
forecasts errors on $\acip$ of order $\sigma_{\acip} = 0.25$. Reference \cite{2019arXiv190400024H} investigated also 
the constraining power of the galaxy distribution alone via the spatial modulation of the BAO peak position that correlated 
CIPs would induce; next-generation galaxy surveys would however return constraints that are comparable to current CMB bounds 
on $\acip$ for uncorrelated CIPs (cf.~Eq.~(\ref{eq:cmbconstraints})). The impact that CIPs (both correlated and uncorrelated) have on spectral distortions of the CMB 
was also recently studied in Ref.~\cite{2018JCAP...08..036H}, but their forecasts show only modest constraining power: $\xi{\acip} \lesssim 10^4$ for cosmic variance-limited experiments.

%**********************
%**********************
\subsection{On the simplifying assumptions of the analysis}
\label{sec:results:discussion}

Before we conclude we discuss briefly some of the simplifying assumptions we have made in our analysis. While these can have an impact on the absolute values of $\sigma_{\fnl}$ and $\sigma_{\acip}$, they should nonetheless have a weaker impact on their relative size, $\sigma_{\acip}/\sigma_{\fnl}$, which is therefore a more trustworthy measure of the constraining power of the galaxy power spectrum. 

For example, although we have considered a few example galaxy sample pairs with different galaxy bias parameters, we have still assumed perfect knowledge of galaxy bias in the constraints from each pair. Results from numerical simulations of galaxy formation can be used to inform priors on galaxy  bias \cite{2018MNRAS.475..676S, 2019arXiv190402070B, 2020JCAP...02..005B}. This is especially important  for constraints on CIPs in light of the marked differences between the constraints using halo  mass- and stellar mass-selected galaxies (cf.~Fig.~\ref{fig:contours_intuition});  
concretely, our simplifying assumption of unmodified stellar-to-halo mass relations by density fluctuations and local PNG  should be revisited and its implications on $b_1(M_*)$ and $b_\phi(M_*)$ analyzed.  Further, it would also be interesting to forecast the constraining power of joint galaxy power  spectrum and bispectrum analysis, and investigate to which extent the incorporation of higher-order statistical information can be helpful in breaking degeneracies with galaxy bias.

Another assumption we have made is that, for each of our four galaxy sample pairs, the higher-mass sample $S_2$ contains $5\%$ of the total number of galaxies, $\bar{n}_g^{S_2} = 0.05\bar{n}_g, \bar{n}_g^{S_1} = \bar{n}_g - \bar{n}_g^{S_2}$; this represents only a rough estimate of the sparsity of samples $S_2$ relative to samples $S_1$ in our chosen example galaxy sample pairs. In general, the larger the ratio $\bar{n}_g^{S_1} / \bar{n}_g^{S_2}$, the more of the constraining power will come from sample $S_1$, and this can impact the constraints on local PNG and CIPs via the efficiency with which the $\fnl-\acip$ degeneracy is broken. Improvements over our simplified treatment here should be done using survey-specific galaxy selection functions to determine the expected observed galaxy number density as various selection cuts are made (in mass, luminosity, etc.). As a test, we have nonetheless explicitly checked that choosing $\bar{n}_g^{S_2} = 0.01\bar{n}_g$ or $\bar{n}_g^{S_2} = 0.10\bar{n}_g$ preserves our main conclusions on the potential of the galaxy power spectrum to constrain CIPs and local PNG simultaneously. 

We have also neglected RSD contributions to the galaxy power spectrum (cf.~Eq.~\eqref{eq:Pgg}). On the largest scales that are relevant for $\fnl$ and $\acip$ constraints, the relevant contribution would come from the Kaiser term \cite{kaiser:1987} that would effectively rescale $b_1 \to b_1 + f\mu^2$, where $f$ is the linear growth rate and $\mu$ is the cosine of the angle between the wavenumber $\vk$ and the line of sight. Approximating $f \sim 1$, this would imply a $\approx 10-30\%$ boost of the contribution from $P_{mm}$ relative to  $P_{m\phi}$ and $P_{m\sigma}$ in the angle-averaged galaxy power spectrum for $b_1 \approx 1-4$. This would in turn reduce the impact of the $\fnl$ and $\acip$ terms and worsen their constraints, though importantly not their ratio $\sigma_{\acip}/\sigma_{\fnl}$. Similar considerations hold to the inclusion of lightcone projection effects.

Again, while going beyond these simplifications is expected to have an impact on the absolute constraints on $\fnl$ and $\acip$, the values of $\sigma_{\acip}/\sigma_{\fnl}$ reported here should remain fairly close to what more robust forecasts and real data analyses will yield.  Specifically, for $\xi = 1$, Fig.~\ref{fig:xipair3} shows that galaxy sample pair 3 can yield $\sigma_{\acip}/\sigma_{\fnl} \approx 1 - 2$. On the other hand, for $\xi = 0$, Figs.~\ref{fig:xipair3} and \ref{fig:xipair1} show that galaxy pairs 3 and 1 could return relative constraints of order $\sigma_{\acip}/\sigma_{\fnl} \approx 5$ and $\sigma_{\acip}/\sigma_{\fnl} \approx 15$, respectively. These values are effectively independent of the survey volume $V_S$ and can be used to roughly forecast constraints on $\sigma_{\acip}$, given existing more robust constraints/forecasts on $\sigma_{\fnl}$. {Importantly, the constraints on $\fnl$ and $\acip$ are affected by the same observational systematics that affect large-scale clustering measurements \cite{2012MNRAS.424..564R, 2013PASP..125..705P, 2013MNRAS.435.1857L, 2019MNRAS.482..453K}, since they are based on the same scale-dependent feature that becomes prominent on the largest scales.}

Specifically, taking $\sigma_{\fnl} = 50$ as a conservative measure of the current constraining power of the galaxy power spectrum on local PNG \cite{slosar/etal:2008, 2014PhRvD..89b3511G, 2014PhRvL.113v1301L, 2014MNRAS.441L..16G, 2015JCAP...05..040H, 2019JCAP...09..010C}, then our results suggest that $\sigma_{\acip} \approx 2\times \sigma_{\fnl} = 100$ may be already possible with existing analysis pipelines; these would constitute the tightest constraints on CIPs to date. Likewise, next-generation galaxy surveys that aim to reach $\sigma_{\fnl} \sim 1$, could probe CIPs with $\sigma_{\acip} \approx 1-2$ for $\xi = 1$ and $\sigma_{\acip} \approx 5-15$ for $\xi = 0$. {This suggests that upcoming galaxy surveys may comfortably improve on the constraining power of cosmic-variance limited CMB data, which has been argued to be $\sigma_A \sim 6$ \cite{2018PhRvD..97b3513H} and $\sigma_A \sim 90$ \cite{2016PhRvD..93d3008M, 2017JCAP...04..014V},\footnote{{In this estimate, we take the claim from Refs.~\cite{2016PhRvD..93d3008M, 2017JCAP...04..014V} that cosmic-variance limited CMB data can improve upon existing constraints by a factor of $\approx 5$, which yields $A \lesssim 450/5 = 90$ from Eq.~(\ref{eq:cmbconstraints}).}}  for $\xi = 1$ and $\xi = 0$, respectively.}

We note that improvements to our analysis here could involve also determining which properties of a large-scale structure survey and constraint analysis are able to return optimal constraints on $\fnl$ and $\acip$ \cite{2017PhRvD..95l3513D}. Some specifications that can have an impact on the resulting constraints include redshift range, galaxy selection variables and expected bias values, number of tracers in the multitracer technique, survey volume vs. galaxy number density, tracers beyond galaxies, etc. These can be explored and exploited to extract the maximum potential out of current and future galaxy survey data. 

%**********************
%**********************
\section{Summary \& Conclusions}
\label{sec:conclusions}

In this paper we have investigated the potential of using the statistics of the late-time galaxy 
distribution to simultaneously constrain local PNG and primordial CIPs. 
A primordial CIP perturbation $\sigma(\vx)$ describes spatial fluctuations 
in the baryon density that are compensated by opposite-sign perturbations in the CDM density to leave the total matter 
density field unperturbed. The CMB data is in general very powerful at constraining the amplitude of both adiabatic 
and generic isocurvature perturbations, but the {\it compensated} nature of CIPs implies that they contribute 
only at second order, which explains why the tightest constraints still allow the amplitude of the power spectrum of CIPs, 
$P_{\sigma\sigma}$, to be over $5$ orders of magnitude larger than that of adiabatic scalar perturbations $P_{\R\R}$ 
(assuming $\xi=0$, i.e. no correlation between $\sigma$ and $\R$). On the other hand, CIPs contribute at leading order in their amplitude 
to the galaxy overdensity, $\delta_g(\vx, z) \supset b_\sigma(z) \sigma(\vx)$ (cf.~Eq.~(\ref{eq:biasexp})), which makes 
it interesting to investigate the sensitivity of galaxy statistics to the statistics of CIPs, and anticipate the constraints that can be achieved with current and future data. That was our main goal. 

We focused on the impact of CIPs on the galaxy power spectrum and parametrized CIPs via (i) 
the relative amplitude $\acip$ of their auto power spectrum to the adiabatic perturbations, 
$P_{\sigma\sigma} = \acip^2P_{\R\R}$, and (ii) a correlation coefficient $\xi = P_{\sigma\R}/\sqrt{P_{\sigma\sigma}P_{\R\R}}$ 
(cf.~Eqs.~(\ref{eq:Psigmasigma_1}) and (\ref{eq:Psigmasigma_2})). To leading order, correlated CIPs ($\xi = 1$) contribute to the galaxy power spectrum as $\propto b_\sigma\acip/k^2$, 
which is exactly the same scale dependence as the contribution from local PNG $\propto b_\phi \fnl / k^2$ (cf.~Fig.~\ref{fig:model} 
and Eqs.~(\ref{eq:Pgg_contributions_1}) - (\ref{eq:Pgg_contributions_7})). Constraints with a single galaxy sample exhibit 
a strong degeneracy along the $\acip = -(3b_\phi/(5b_\sigma))\fnl$ direction, effectively resulting in very poor constraints 
in either one of the parameters. Here, we relied on the galaxy multitracer technique to seek pairs of galaxy 
samples with different galaxy bias parameters $b_\phi$ and $b_\sigma$, as a way to break the degeneracy in $\fnl-\acip$ space. 
We made use of known analytical formulae to calculate these bias parameters as a function of total host halo mass, 
and of a simplified treatment of the stellar-to-halo mass relation to describe them in terms of galaxy stellar mass 
(cf.~Fig.~\ref{fig:bias_intuition} and Sec.~\ref{sec:results:bias}). 

Our main conclusions can be summarized as follows:

\begin{itemize}%[leftmargin=*]

\item For correlated CIPs ($\xi = 1$), galaxy sample pairs with stellar mass-selected samples can break the $\fnl-\acip$ degeneracy 
more efficiently than selecting by halo mass, and can return tighter constraints in general
(cf.~Fig.~\ref{fig:contours_intuition}). This can be traced back to the stronger dependence on mass 
and redshift of the ratio $b_\phi/b_\sigma$ for stellar mass selection (for sizable $b_\phi$ and $b_\sigma$; cf.~Fig.~\ref{fig:bias_intuition}).

\item The contribution from uncorrelated CIPs ($\xi = 0$) scales as $\acip^2/k^4$ and it is thus not degenerate with 
$\fnl$ (cf.~$\xi = 0$ vs. $\xi = 1$ contours in Figs.~\ref{fig:xipair3} and \ref{fig:xipair1}); this reduces the 
importance of galaxy selection and bias to break the degeneracy present in the $\xi = 1$ case. 
For example, galaxy sample pairs which return weak constraints on both $\fnl$ and $\acip$ for $\xi=1$ 
can yield tighter constraints if $\xi = 0$ (cf.~Fig.~\ref{fig:xipair1}).

\item Taking external prior information on $\fnl$ into account is another way to break the degeneracy for correlated CIPs 
($\xi = 1$), and results in constraints on $\acip$ that are generically tighter and less critically dependent on galaxy selection and bias (cf.~Fig.~\ref{fig:fnlprior}).

\item Our results indicate that galaxy power spectrum constraints on $\acip$ relative to $\fnl$ have the potential to be of order $\sigma_{\acip}/\sigma_{\fnl} \approx 1-2$ for $\xi = 1$ and $\sigma_{\acip}/\sigma_{\fnl} \approx 5$ for $\xi = 0$ (cf.~Sec.~\ref{sec:results:comparison}). {This is better than what cosmic-variance limited CMB data can achieve.}

\end{itemize}

Our findings provide strong motivation to place constraints on $\acip$ from existing galaxy survey data, which can directly build on analysis pipelines {(including control of systematics)} used already for constraints on local PNG. The tightest current constraints on $\fnl$ from the galaxy power spectrum are of order $\sigma_{\fnl} \sim 50$ \cite{slosar/etal:2008, 2013MNRAS.428.1116R, 2014PhRvD..89b3511G, 2014PhRvL.113v1301L, 2014MNRAS.441L..16G, 2015JCAP...05..040H, 2019JCAP...09..010C}. Taking the numerical values of $\sigma_{\acip}/\sigma_{\fnl}$ obtained from our example galaxy sample pairs at face value indicates that constraints of order $\sigma_{\acip} \sim 50-100$ for $\xi = 1$ and $\sigma_{\acip} \sim 250$ for $\xi = 0$ might be possible with existing data already. {For $\xi = 0$, these} can be compared with the current tightest bounds from the CMB, $\acip \lesssim 450$, which could therefore already be improved upon with existing galaxy samples. Given the potential strong degeneracies between $\fnl$ and $\acip$, which had not been recognized previously, the impact of relaxing the $\acip = 0$ assumption currently made in essentially all constraints and forecasts for $\fnl$ from galaxy statistics should also be investigated.

{Finally, it is interesting to inspect the implications of our findings to the case of the curvaton model, which is a popular model of inflation with two fields that can generate correlated CIPs \cite{1997PhRvD..56..535L, 2000PhRvD..62d3504L, 2003PhRvD..67l3513G, 2003PhRvD..67b3503L, 2006RvMP...78..537B, 2006PhRvD..74j3003S, 2015PhRvD..92f3018H}. In the curvaton model, different amounts of CIPs can be generated depending on whether baryons and CDM are produced before, by or after the decay of the curvaton field \cite{2003PhRvD..67l3513G, 2015PhRvD..92f3018H}. The two most observationally interesting scenarios with potentially detectable correlated CIPs yield $A \approx 16$ and $A\approx -3$. Our results show that both are expected to be well within the detection ability of future galaxy surveys for $\xi =1$: $\sigma_{\acip} \approx 1-2 \times \sigma_{\fnl} \approx 1-2$ for $\sigma_{\rm fnl} \sim 1$. Note that these curvaton scenarios can also produce non-negligible amounts of local PNG with $|\fnl| \approx 6$ and $|\fnl| \approx 1$ for $A \approx 16$ and $A \approx -3$, respectively \cite{2006PhRvD..74j3003S}. This strengthens the case for simultaneously constraining both $A$ and $\fnl$ using galaxy clustering data.} 

%**********************
%**********************
\acknowledgments

AB, GC and FS acknowledge support from the Starting Grant (ERC-2015-STG 678652) ``GrInflaGal'' from the European Research Council.

%**********************
%**********************
\appendix

%**********************
%**********************
\section{Covariance matrix of the bin-averaged galaxy power spectrum}\label{app:estimator}

In this appendix we display a few steps of the derivation of the covariance matrix 
of the galaxy power spectrum estimator used in the main body of the paper. 

Repeating some equations here for self-containedness, our data vector is composed of an estimator for 
the auto power spectrum of the galaxy sample $A$, $\smash{\hat{P}_{gg}^{S_1S_1}}$, the cross-spectrum between galaxy 
samples $S_1$ and $S_2$, $\smash{\hat{P}_{gg}^{S_1S_2}}$ and the auto power spectrum of galaxy sample $S_2$, $\smash{\hat{P}_{gg}^{S_2S_2}}$:
\bq\label{eq:datavector_app}
{\bf D}(k) = \Big\{\hat{P}_{gg}^{S_1S_1}, \hat{P}_{gg}^{S_1S_2}, \hat{P}_{gg}^{S_2S_2}\Big\}\,\,,
\eq
with 
\bq\label{eq:angavePgg-app}
\hat{P}_{gg}^{S_1S_1}(k) &=& \frac{1}{V_SV_k} \int_{k} {\rm d}^3\vk'\ {\delta}_{g}^{S_1}(\vk'){\delta}_g^{S_1}(-\vk')\,\,, \\
\hat{P}_{gg}^{S_1S_2}(k) &=& \frac{1}{V_SV_k} \int_{k} {\rm d}^3\vk'\ {\delta}_{g}^{S_1}(\vk'){\delta}_g^{S_2}(-\vk')\,\,, \\
\hat{P}_{gg}^{S_2S_2}(k) &=& \frac{1}{V_SV_k} \int_{k} {\rm d}^3\vk'\ {\delta}_{g}^{S_2}(\vk'){\delta}_g^{S_2}(-\vk')\,\,.
\eq
These estimators are unbiased as, e.g., 
\bq
\langle \hat{P}_{gg}^{S_1S_1}(k) \rangle &=& \frac{1}{V_SV_k} \int_{k} {\rm d}^3\vk' \langle{\delta}_{g}^{S_1}(\vk'){\delta}_g^{S_1}(-\vk')\rangle \\
&=& \frac{1}{V_SV_k} \int_{k} {\rm d}^3\vk' (2\pi)^3 P_{gg}^{S_1S_1}(k') \delta_D(\vec{0}) \\ 
&=& \frac{1}{V_k} \int_{k} {\rm d}^3\vk' P_{gg}^{S_1S_1}(k') \approx P_{gg}^{S_1S_1}(k)\,\,,
\eq
where the last approximation holds for sufficiently small bin widths for it to be a good approximation to take the power spectrum out of the integral and we have also used $\smash{\delta_D(\vec{0}) \equiv V_S/(2\pi)^3}$; similar steps hold for $\smash{\hat{P}_{gg}^{S_1S_2}(k)}$ and $\smash{\hat{P}_{gg}^{S_2S_2}(k)}$.

The covariance matrix of the data vector ${\bf Cov} \equiv {\bf Cov}(k_1, k_2) = \big<{\bf D}(k_1){\bf D}(k_2)\big> - \big<{\bf D}(k_1)\big>\big<{\bf D}(k_2)\big>$ can be written as 
\begin{equation}
{\bf Cov} =
\begin{pmatrix}
{\rm Cov}\left[\hat{P}_{gg}^{S_1S_1}(k_1), \hat{P}_{gg}^{S_1S_1}(k_2)\right] & {\rm Cov}\left[\hat{P}_{gg}^{S_1S_2}(k_1), \hat{P}_{gg}^{S_1S_1}(k_2)\right] & {\rm Cov}\left[\hat{P}_{gg}^{S_2S_2}(k_1), \hat{P}_{gg}^{S_1S_1}(k_2)\right] \\[1.5ex]
\cdots & {\rm Cov}\left[\hat{P}_{gg}^{S_1S_2}(k_1), \hat{P}_{gg}^{S_1S_2}(k_2)\right] & {\rm Cov}\left[\hat{P}_{gg}^{S_2S_2}(k_1), \hat{P}_{gg}^{S_1S_2}(k_2)\right] \\[1.5ex]
\cdots & \cdots & {\rm Cov}\left[\hat{P}_{gg}^{S_2S_2}(k_1), \hat{P}_{gg}^{S_2S_2}(k_2)\right]
\end{pmatrix}
\,\,,
\end{equation}
where each block represents the covariance of the spectrum estimators. We skip writing the entries marked with ``$\cdots$''; 
the covariance is a symmetric matrix. Let us consider explicitly the ${\rm Cov}\left[\hat{P}_{gg}^{S_1S_2}(k_1), \hat{P}_{gg}^{S_1S_2}(k_2)\right]$ term:
\begin{equation}\begin{split}
{\rm Cov}\left[\hat{P}_{gg}^{S_1S_2}(k_1), \hat{P}_{gg}^{S_1S_2}(k_2)\right] &= \frac{1}{V_S^2V_{k_1}V_{k_2}} \int_{k_1} {\rm d}^3\vk \int_{k_2} {\rm d}^3\vk' \langle{\delta}_g^{S_1}(\vk){\delta}_g^{S_2}(-\vk){\delta}_g^{S_1}(\vk'){\delta}_g^{S_2}(-\vk')\rangle \\ 
&\;\;\;\; - P_{gg}^{S_1S_2}(k_1)P_{gg}^{S_1S_2}(k_2)\,\,.
\end{split}\end{equation}
Applying Wick's theorem to the four-point function yields three terms proportional to two two-point functions and a term given by the connected four-point function. One of the two-point function terms cancels exactly with the $P_{gg}^{S_1S_2}(k_1)P_{gg}^{S_1S_2}(k_2)$ term, while the contribution from the four-point function is negligible on the large-scales (low $k$) we are interested in this paper. The derivation thus continues as
\bq
{\rm Cov}\left[\hat{P}_{gg}^{S_1S_2}(k_1), \hat{P}_{gg}^{S_1S_2}(k_2)\right] &=& \frac{1}{V_S^2V_{k_1}V_{k_2}} \int_{k_1} {\rm d}^3\vk \int_{k_2} {\rm d}^3\vk' \langle{\delta}_g^{S_1}(\vk){\delta}_g^{S_1}(\vk')\rangle\langle{\delta}_g^{S_2}(-\vk){\delta}_g^{S_2}(-\vk')\rangle \nonumber \\
&+&\frac{1}{V_S^2V_{k_1}V_{k_2}} \int_{k_1} {\rm d}^3\vk \int_{k_2} {\rm d}^3\vk' \langle{\delta}_g^{S_1}(\vk){\delta}_g^{S_2}(-\vk')\rangle\langle{\delta}_g^{S_2}(-\vk){\delta}_g^{S_1}(\vk')\rangle \nonumber \\
&=& \frac{1}{V_S^2V_{k_1}V_{k_2}} \int_{k_1} {\rm d}^3\vk \int_{k_2} {\rm d}^3\vk' (2\pi)^6 P_{gg}^{S_1S_1}(k)P_{gg}^{S_2S_2}(k) \delta_D(\vk + \vk') \delta_D(-\vk - \vk') \nonumber \\
&+& \frac{1}{V_S^2V_{k_1}V_{k_2}} \int_{k_1} {\rm d}^3\vk \int_{k_2} {\rm d}^3\vk' (2\pi)^6 [P_{gg}^{S_1S_2}(k)]^2 \delta_D(\vk - \vk') \delta_D(-\vk + \vk') \nonumber \\
&=& \frac{(2\pi)^6 \delta_{k_1k_2}}{V_S^2V_{k_1}V_{k_2}} \int_{k_1} {\rm d}^3\vk\ \left[P_{gg}^{S_1S_1}(k)P_{gg}^{S_2S_2}(k) + [P_{gg}^{S_1S_2}(k)]^2\right]\delta_D(\vec{0})\,\,,
\eq
where the Kronecker symbol reflects that the integral over ${\rm d}^3\vk'$ is only non-zero if both averages are for the same $k$-bin. Further keeping with the assumption of sufficiently small bin-widths for one to skip explicitly the bin average we have
\bq
{\rm Cov}\left[\hat{P}_{gg}^{S_1S_2}(k_1), \hat{P}_{gg}^{S_1S_2}(k_2)\right] = \frac{(2\pi)^3 \delta_{k_1k_2}}{V_SV_{k_1}} \left[P_{gg}^{S_1S_1}(k_1)P_{gg}^{S_2S_2}(k_1) + [P_{gg}^{S_1S_2}(k_1)]^2\right]\,\,,
\eq
which matches the corresponding entry in the covariance matrix of Eq.~(\ref{eq:covariance}); the remaining entries are obtained straightforwardly by following analogous steps. The auto power spectra $P_{gg}^{S_1S_1}$ and $P_{gg}^{S_2S_2}$ is given by Eq.~(\ref{eq:Pgg}). The cross-spectrum of the galaxy samples $S_1$ and $S_2$ is given by 
\bq\label{eq:Pggab}
P_{gg}^{S_1S_2}(k, z) &=& \Big[\frac{9}{25}b_1^{S_1} b_1^{S_2} \mathcal{M}^2(k, z) 
+ \frac{9}{25}\fnl\left(b_1^{S_1}b_\phi^{S_2} + b_1^{S_2}b_\phi^{S_1}\right)\mathcal{M}(k, z) \nonumber \\
&+& \frac{3}{5}\xi \acip \left(b_1^{S_1}b_\sigma^{S_2}+b_1^{S_2}b_\sigma^{S_1}\right)\mathcal{M}(k,z) 
+ \frac{9}{25}\fnl^2b_\phi^{S_1}b_\phi^{S_2} \nonumber \\
&+& \frac{3}{5}\fnl\xi\acip\left(b_\phi^{S_1}b_\sigma^{S_2} + b_\phi^{S_2}b_\sigma^{S_1}\right) 
+ b_\sigma^{S_1}b_\sigma^{S_2}\acip^2\Big]P_{\R\R}(k) \,, \nonumber \\
\eq
where we neglect the cross-spectrum of the noise fields $\epsilon_{S_1}$ and $\epsilon_{S_2}$ \cite{hamaus/etal:2010}.

%**************************************
%**************************************
\bibliographystyle{utphys}
\bibliography{REFS}

%**************************************
%**************************************
%**************************************
\end{document}